\documentclass{aa}  

\usepackage{graphicx}
\usepackage{txfonts}
\usepackage{comment}
\usepackage[colorlinks,allcolors=blue]{hyperref}

\newcommand{\ha}          {H$\alpha$}
\newcommand{\hb}          {H$\beta$}
\newcommand{\wha}          {$W_{\rm H\alpha}$}

\newcommand{\kmpers}      {\mbox{\rm km~s$^{-1}$}}
\newcommand{\kms} {\kmpers}

\newcommand{\aco}         {\mbox{$\alpha_{\rm CO(1-0)}$}}

\newcommand{\fmol}         {\mbox{$f_{\rm mol}$}}

\newcommand{\iedge}{iEDGE}

\defcitealias{kalinova2021}{K21}
\defcitealias{colombo2020}{C20}

\begin{document}

   \title{The EDGE-CALIFA Survey: An integral field unit-based integrated molecular gas database for galaxy evolution studies in the Local Universe}

   \subtitle{}

   \author{D. Colombo\inst{1,2}\thanks{dcolombo@uni-bonn.de},
          V. Kalinova\inst{2},
          Z. Bazzi\inst{1},
          S.F. Sanchez\inst{3,4},
          A. D. Bolatto\inst{5},
          T. Wong\inst{6},
          V. Villanueva\inst{5,7},\\
          N. Mudivarthi\inst{1},
          E. Rosolowsky\inst{8},
          A. Wei\ss\inst{2},
          K. D. French\inst{6},
          A. Leroy\inst{9},
          J. Barrera-Ballesteros\inst{3},
          Y. Garay-Solis\inst{3},\\ 
          F. Bigiel\inst{1},
          A. Tripathi\inst{6},
          B. Rodriguez\inst{1}
          }

   \institute{Argelander-Institut f\"ur Astronomie, University of Bonn, Auf dem H\"ugel 71, 53121 Bonn, Germany
   \and
   Max-Planck-Institut f\"ur Radioastronomie, Auf dem H\"ugel 69, 53121 Bonn, Germany
   \and
   Universidad Nacional Aut\'onoma de M\'exico, Instituto de Astronom\'\i a, AP 106, Ensenada 22800, BC, M\'exico
   \and
   Instituto de Astrof\'\i sica de Canarias, V\'\i a L\'actea s/n, 38205, La Laguna, Tenerife, Spain
   \and
   Department of Astronomy, University of Maryland, College Park, MD 20742, USA
   \and
   Department of Astronomy, University of Illinois, Urbana, IL 61801, USA
   \and
   Departamento de Astronom{\'i}a, Universidad de Concepci{\'o}n, Barrio Universitario, Concepci{\'o}n, Chile
   \and
   Department of Physics, University of Alberta, 4-181 CCIS, Edmonton, AB T6G 2E1, Canada
   \and
   Department of Astronomy, The Ohio State University, 140 West 18$^{\rm th}$ Avenue, Columbus, OH 43210, USA}

   \date{Received XXX; accepted XXX}

  \abstract{Studying galaxy evolution requires knowledge not only of the stellar properties, but also of the interstellar medium (in particular the molecular phase) out of which stars form, using a statistically significant and unbiased sample of galaxies. To this end, we introduce here the integrated Extragalactic Database for Galaxy Evolution (\iedge), a collection of integrated stellar and nebular emission lines, and molecular gas properties from 643 galaxies in the local Universe. These galaxies are drawn from the CALIFA  datasets, and are followed up in CO lines by the APEX, CARMA, and ACA telescopes. As this database is assembled from data coming from a heterogeneous set of telescopes (including IFU optical data and single-dish and interferometric CO data), we adopted a series of techniques (tapering, spatial and spectral smoothing, and aperture correction) to homogenise the data. Due to the application of these techniques, the database contains measurements from the inner regions of the galaxies and for the full galaxy extent. We used the database to study the fundamental star formation relationships between star formation rate (SFR), stellar mass ($M_*$), and molecular gas mass ($M_{\rm mol}$) across galaxies with different morphologies. We observed that the diagrams defined by these quantities are bi-modal, with early-type passive objects well separated from spiral star-forming galaxies. Additionally, while the molecular gas fraction ($f_{\rm mol}=M_{\rm mol}/M_*$) decreases homogeneously across these two types of galaxies, the star formation efficiency (SFE=SFR/$M_{\rm mol}$) in the inner regions of passive galaxies is almost two orders of magnitude lower compared to the global values. This indicates that   inside-out quenching requires not only low $f_{\rm mol}$, but also strongly reduced SFE in the galactic centres.}


   \titlerunning{integrated Extragalactic Database for Galaxy Evolution}

   \authorrunning{D. Colombo, V. Kalinova, S. Sanchez et al.}

   \maketitle

\section{Introduction}\label{S:introduction}
Star formation is one of the main processes governing the evolution of galaxies, with stars forming from the coldest phase of the interstellar medium (ISM), the molecular gas \citep{mckee07}. Molecular gas is primarily composed of molecular hydrogen (H$_2$). However, H$_2$ lacks a permanent dipole moment, and its quadrupole rotational transitions require very high temperatures to be excited ($\gtrsim 500$~K; e.g. \citealt{draine2011}), while its ultraviolet transitions, which can probe colder H$_2$, are limited to diffuse gas along specific sightlines and are largely impractical for characterising the bulk of the molecular ISM in galaxies. Consequently, rotational transitions of the second most abundant molecule, carbon monoxide ($^{12}$C$^{16}$O), along with its isotopologues, are widely used as the primary tracers of cold molecular gas. For integrated measurements, additional methods based on the total dust mass, combined with an assumed dust-to-gas mass ratio, can also provide estimates of the total molecular gas content \citep[e.g.][]{leroy2011, sandstrom2013}. However, such methods rely heavily on empirical calibrations and are often less feasible for spatially resolved studies at the resolution of nearby galaxy surveys (but see \citealt{barrera-ballesteros2015,piotrowska2020}). This means  that the CO tracer remains the most direct and widely adopted approach for estimating the molecular gas content.

\subsection{From CO-integrated surveys to study galaxy evolution ...}

From the early detection of the CO molecule in the Orion nebula \citep{wilson1970}, the observations of CO have been routinely used to infer the molecular gas amount, also in extragalactic objects \citep{verter1985}. While mapping the CO emission offers a complete view of the distribution and the kinematics of the molecular gas in galaxies, integrated measurements where all the CO flux information from a galaxy is collected into a single emission line spectrum provide a way to obtain significant population statistics with relatively simple observational set-ups  and short exposure times. For this reason, integrated CO surveys have been the main tool for calibrating global scaling relations and understanding the properties of molecular gas across different physical conditions. As blind CO surveys cannot be easily performed with current and past facilities, CO surveys have been generally biased towards certain galaxy types or parameter ranges, such as spiral \citep[e.g.][]{braine1993,sage1993,sanders_mirabel1985,young1995}, early-type \citep[e.g.][]{thronson1989,wiklind_henkel1989,lees1991,wang1992,wiklind1995,combes2007,welch2010,young2011}, and cluster galaxies \citep[e.g.][]{kenney1988,boselli1997}; high-mass \citep[e.g.][]{saintonge2011,saintonge2017,keenan2024} and low-mass galaxies \citep[e.g.][]{bothwell2014,cicone2017}; and also distant \citep[e.g.][]{tacconi2010,tacconi2013}, isolated \citep[e.g.][]{rodriguez2024}, and active objects \citep[see e.g.][and references therein]{koss2021}.

Given their ability to span large ranges of parameter space, integrated surveys are particularly useful in order to constrain the physics driving galaxy evolution. This boils down to understanding the galaxy distribution across the colour-magnitude diagram or the star formation rate--stellar mass (SFR--$M_*$) diagram. Galaxies in the local Universe tend to populate two prominent regions in these diagrams \citep[e.g.][]{renzini_peng2015}: the star formation main sequence (SFMS; e.g. \citealt{brinchmann2004,cano_diaz2016}) or blue cloud where typically star-forming, spiral galaxies reside, and the red or retired sequence, dominated mostly by early-type passive objects. The region between the two, called the green valley \citep{salim2007}, is sparsely populated and comprises galaxies that are slowly decreasing their star formation activity (i.e. they are quenching). From the point of view of the integrated CO surveys, studying galaxy evolution means understanding the relation between the molecular gas ($M_{\rm mol}$) availability across the galaxies (parametrised as the molecular gas fraction, $f_{\rm mol}=M_{\rm mol}/M_*$) and  the rate at which this gas is converted into stars (parametrised as the star formation efficiency, SFE=SFR/$M_{\rm mol}$, or its inverse, the depletion time, $\tau_{\rm dep}$=1/SFE) in the different regions of the SFR-$M_*$ diagram,  and how varying these quantities moves the galaxies across the diagram. For example, the SFMS is not a straight line across the diagram, but presents scatter and a possible flattening at high masses \citep[e.g.][]{saintonge_catinella2022}. This might be due to the increment of bulged galaxies at the high mass end \citep[e.g.][]{bluck2014,cook2020}, which are generally massive but not star-forming, or by a lower gas availability or reduced SFE in high-mass galaxies that causes a lower specific star formation rate (sSFR=SFR/$M_*$). Bars can also play a role (\citealt{hogarth2024,scaloni2024}, but see \citealt{diaz_garcia2021}) by funnelling gas in the centre, which causes a nuclear starburst and rapid gas consumption, preventing further star formation. Some of these phenomena might even move the galaxies away from the SFMS, further down to the green valley and eventually to the retired region. It is generally accepted that galaxies below the main sequence are gas poor \citep[e.g.][]{young2011,davis2019}, indicating that quenching galaxies need to get rid of their cold gas budget to fully retire \citep[e.g.][]{tacconi2013,genzel2015,tacconi2018,michalowski2024}. While the powerful outflows driven by active galactic nuclei (AGNs; e.g. \citealt{cicone2014,costa_souza2024}) are often invoked as the main mechanisms (in high-mass galaxies) to remove molecular gas \citep[e.g.][]{piotrowska2022}, some integrated surveys have shown that active galaxies have (globally) even higher molecular gas masses than their non-active counterparts (with the same stellar mass; \citealt{saintonge2017,rosario2018,koss2021}). Other studies have indicated that the reduction of the molecular gas fraction is not enough to explain the quenching of galaxies, but SFE needs to be reduced as well \citep[e.g.][]{saintonge2016,colombo2020,michalowski2024}. This can happen due to dynamical effects, from the increased gravitational potential from the bulge to shears and streaming motions, that stabilise the gaseous disc, preventing fragmentation and star formation \citep[e.g.][]{davis2014}.

\subsection{... to IFU-spatially resolved follow-ups.}

In recent years, some integrated CO surveys \citep[e.g.][]{young2011,saintonge2018,colombo2020,dominguez_gomez2022,wylezalek2022} have focused on the follow-up of integral field unit (IFU) optical surveys. These  surveys spatially and spectrally resolve the optical emission across galaxies and provide extensive amounts of information from the stellar continuum (such as stellar mass surface densities, velocities, ages, metallicities, and star formation histories) and emission lines (to estimate star formation rates, gas kinematics, and gas ionisation sources). The combination of this information with measurements from CO emission data (namely molecular gas masses and CO velocities) can provide valuable insights into galaxy evolution, such as understanding why certain galaxies exhibit unusually high or low star formation rates relative to the local average. 

Parallel to integrated studies, a series of spatially resolved CO-based surveys in the local Universe (typically at $z<0.03$) has been conducted, specifically designed to address key questions in galaxy evolution by building on IFU datasets. These resolved works refined the results from global studies, and allowed us to trace local variations of SFE and \fmol\ and the morphological, energetical, and kinematical features of the galaxies that potentially drive those variations. In particular the Extragalactic Database for Galaxy Evolution collaboration have followed up a sub-set of 177 galaxies observed in $^{12}$CO(1-0) by the Combined Array for Research in Millimeter-wave Astronomy (CARMA) telescope \citep{bolatto2017} plus 60 galaxies observed in $^{12}$CO(2-1) by the Atacama Compact Array (ACA) telescope \citep{villanueva2024}, drawn from the Calar Alto Legacy Integral Field Area IFU survey mother sample \citep{sanchez2012,sanchez2016}. Similarly, the ALMA-MaNGA QUEnching and STar formation (ALMaQUEST, \citealt{lin2020}) project followed up in $^{12}$CO(1-0) with the Atacama Large Millimeter/submillimeter Array (ALMA) a sample of 46 galaxies drawn from the Mapping Nearby Galaxies at Apache Point Observatory (MaNGA, \citealt{bundy2015}) IFU survey mother sample. Both projects still broadly concentrate on main sequence galaxies, with a few objects selected from the green valley, starburst, and merger regimes. Earlier, the ATLAS$^{\rm 3D}$ IFU project \citep{cappellari2011} collected interferometric $^{12}$CO(1-0) maps for 39 early-type galaxies \citep{alatalo2013}.

Several works based on those surveys are focused on the resolved relations between SFR, $M_*$, and $M_{\rm mol}$ (in their area-averaged forms), finding that the three quantities describe a tight relation in space \citep{lin2019,sanchez2021}, which hold from kiloparsec- to global-scales and that the three relations between SFR, $M_*$, and $M_{\rm mol}$ are just projections of the three-dimensional relation on two-dimensional planes \citep{sanchez2020}. Additionally, those resolved relations vary from galaxy to galaxy \citep{ellison2021a}, between galaxies \citep{lin2022}, and for galaxies in the green valley \citep{lin2022} or in the starburst regime \citep{ellison2020}. Through these studies it appears that the SFR-$M_*$ relation might be just a bi-product of the relation between  SFR with $M_{\rm mol}$ and $M_{\rm mol}$ with $M_*$ \citep{ellison2021a,baker2022}; and that a `hidden' parameter, the environmental pressure, self-regulates the star formation (and possibly also the SFE) on kiloparsec-scales \citep{barrera-ballesteros2020,sanchez2021,ellison2024}, through supernovae feedback \citep{barrera-ballesteros2020} in star-forming regions or galaxies, and through other processes (such as magnetic fields, cosmic rays, or turbulence-driven inflows) in retired regions \citep{barrera-ballesteros2020} or starburst systems \citep{ellison2024}.

The role of the SFE and \fmol\ in driving the quenching in specific galactic regions has been extensively explored using the EDGE-CALIFA and ALMaQUEST datasets. While SFE and \fmol\ remain relatively constant along the spaxels of main sequence galaxies, they exhibit significant declines in green valley galaxies and retired regions \citep{lin2022,villanueva2024}. The central drop in SFE is identified as a major factor in the quenching of central regions, whereas the quenching in galaxy discs results from a combination of lower SFE and reduced molecular gas fractions \citep{lin2022,pan2024}. Nevertheless, galaxies have also shown higher or lower SFEs in their centre, which are not ascribable to uncertainties in the CO-to-H$_2$ conversion factors \citep{utomo2017}.

Furthermore, these surveys started to tackle the causes behind the quenching of some galactic regions and not only the mechanisms, which appear to vary based on galaxy environment and morphology. Several works based on those datasets found that the dynamical effects associated with bars have strong influences on the star formation and molecular gas distributions in the galaxies. Bar (and also spiral arm, \citealt{yu2022}) dynamics increase the star formation and molecular gas concentration in the centre of galaxies \citep{chown2019}. However, the reduction of the SFE in the centre of galaxies appears to be associated mostly with the presence of radial inflows across the bars rather than with the bars themselves \citep{hogarth2024}. Environmental processes such as gas stripping in Virgo cluster galaxies and minor mergers influence whether molecular gas originates internally or externally \citep{davis2011,davis2013,alatalo2013}, although the star formation in mergers themselves (at any stage) do not appear to be dominated by mainly SFE or \fmol\ variations \citep{thorp2022}. AGN feedback, which is often invoked by simulated work to explain the quenching in high-mass galaxies \citep[e.g.][]{dubois2013,yuan2018,appleby2020,donnari2021,ni2024}, appears to have a limited observational effect on central gas distributions (but see \citealt{ellison2021c}), although it may contribute to long-term quenching by restricting gas accretion \citep{yu2022}. Furthermore, galaxy dynamics, including stabilising shear and morphological effects, can also suppress SFE, particularly in dynamically evolved systems \citep{utomo2017,colombo2018,villanueva2021}. 

\subsection{The need for a IFU-based integrated molecular gas database}\label{SS:intro_iedge}

Those IFU-based projects have provided key insights into the spatially resolved interplay between molecular gas and star formation. However, they are all still biased towards particular galaxy samples. To overcome this limitation, in this paper we introduce iEDGE (the integrated Extragalactic Database for Galaxy Evolution) to expand the available molecular gas information for CALIFA galaxies by providing integrated CO measurements for a sample of 643 CALIFA galaxies, covering a wide and uniform range of star formation activity, from the star-forming main sequence to the green valley and fully quenched galaxy populations. This database builds on the work presented by \citet[][hereafter \citetalias{colombo2020}]{colombo2020}, who defined a dataset of 472 galaxies observed in $^{12}$CO(1-0) by CARMA (177 galaxies) and in $^{12}$CO(2-1) by the Atacama Pathfinder EXperiment (APEX, 296 galaxies). The conclusions of \citetalias{colombo2020} align with those of EDGE-CALIFA and ALMaQUEST, highlighting the key role of reduced central SFE in driving galaxy quenching. However, this result was obtained using a larger sample of quenching galaxies, including systems located significantly further below the SFMS than those covered by EDGE-CALIFA and ALMaQUEST. Moreover, this work demonstrated the unique advantage of combining integrated CO measurements with IFU data, as the latter enables direct identification of the star formation and quenching state in galaxy centres via spatially resolved H$\alpha$ equivalent width—an analysis not feasible with single-fibre or long-slit spectroscopy. Here, we include new CO measurements for 172 CALIFA objects in the database, also considering observations conducted by the EDGE collaboration with the ACA.

While the iEDGE does not offer spatially resolved molecular gas maps, its combination with the spatially resolved optical spectroscopic data from CALIFA provides a powerful framework for characterising the molecular gas content, star formation efficiency, and the physical processes regulating star formation and quenching across a diverse galaxy population and for a statistically significant sample. The combination of iEDGE’s integrated CO measurements with the spatially resolved spectroscopic data from CALIFA enables, for example, a robust determination of key parameters necessary to estimate molecular gas masses, including constraints on the CO(2-1)/CO(1-0) line ratio and the CO-to-H$_2$ conversion factor, using state-of-the-art models. More importantly, this combination will allow a comprehensive analysis of quenching mechanisms within a unified framework, leveraging CALIFA’s spatially resolved nebular emission line diagnostics, stellar population gradients, and kinematics.

\subsection{Paper structure}\label{SS:intro_paper}

The paper is structured as follows. Datasets and surveys used for the \iedge\ construction are described in Section~\ref{S:data}. Methods to obtain a homogeneous database that includes measurements from the inner regions of the galaxies and globally integrated quantities, together with quantity derivation, and comparison between CO fluxes (obtained by different telescopes) are illustrated in Section~\ref{S:methods}. Results from the \iedge\ are collected in Section~\ref{S:results}, where we expose the basic statistics of the database, such as stellar mass and star formation rate distributions, with respect to the mother CALIFA sample; scaling relations between galaxy properties, and comparisons between global and inner galaxy measurements. The conclusions of our work are exposed in Section~\ref{S:conclusion}.

\section{Data}\label{S:data}
The EDGE collaboration \citep{bolatto2017} focuses on connecting optical IFU datasets (principally from CALIFA; \citealt{sanchez2012,sanchez2016}) to data from a variety of sub-millimetre/millimetre telescopes. Here, we present a homogenised database of $^{12}$CO observations of 643 galaxies in the CALIFA dataset. Most of these measurements come from APEX observations, but we supplement these measurements with data from the CARMA and ACA telescopes. In the following, CO observation sensitivities are expressed in terms of molecular gas mass using the standard Milky Way CO(1-0)-to-H$_2$ conversion factor, $\alpha_{\rm CO(1-0)}=4.35$ M$_{\odot}$\,(pc$^2$\,K\,km\,s$^{-1}$)$^{-1}$, and a CO(2-1)/CO(1-0) flux ratio of $R_{21}=0.7$. These values are adopted solely for the purpose of expressing observational limits and do not imply that the sample galaxies necessarily share these Milky Way-like properties. For simplicity, we will define the CO line luminosity, $L'_{\rm CO}$, measured in K\,km\,s$^{-1}$\,pc$^2$, as $L_{\rm CO}$.

\begin{figure*}
    \centering
    \includegraphics[width = 0.85\paperwidth, keepaspectratio]{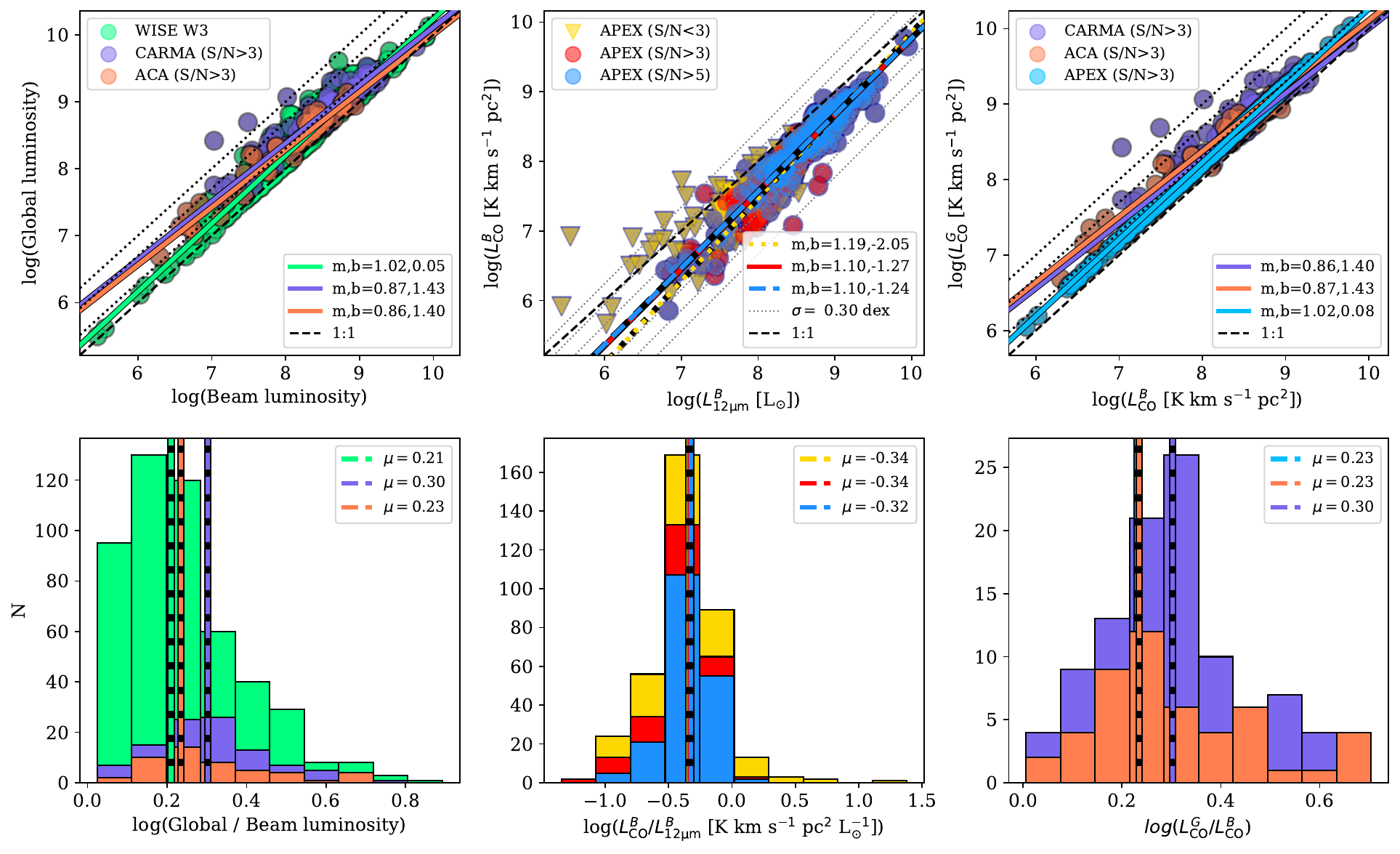}
    \caption{Aperture correction for APEX observation-related tests and measurements. \emph{Upper left:} Global versus beam luminosities calculated for WISE W3 (12\,$\mu$m; green), ACA (red), and CARMA (blue) detected galaxies. The dotted lines indicate the 2:1, 5:1, and 10:1 loci, respectively. \emph{Upper middle:} CO(2-1) luminosities from APEX measurements ($L_{\rm CO}^{B}$) vs WISE W3 (12\,$\mu$m) luminosities ($L_{\rm 12\mu m}^{B}$) calculated within the APEX beam. The blue circles show APEX galaxies detected at $5\sigma_{\rm RMS}$, the red circles indicate APEX galaxies detected at $3\sigma_{\rm RMS}$, while the yellow triangles show non-detected galaxies (with S/N<3) for which an upper limit of the CO luminosity is used. \emph{Upper right:} Global CO luminosities ($L_{\rm CO}^{G}$) vs beam CO luminosities ($L_{\rm CO}^{B}$) for ACA (red), CARMA (blue), and APEX (cyan). For the last, global luminosities are inferred from the aperture correction given by equation~\ref{E:aper_corr}. In the upper panels, fits to the data obtained through the \emph{linmix} method (see Section~\ref{E:aper_corr} for further details) are indicated with full lines using the same colours as the samples. Fit parameters, such as slope ($m$) and intercept ($b$), are also reported in the legend. Additionally, the 1:1 locus is indicated by a dashed black line. In the lower panels, the histograms for the ratios related to the quantities and the samples in the corresponding upper panels are shown, together with the median of the distributions indicated with dashed lines. We can conclude that the global CO luminosities inferred from APEX data using the aperture correction procedure that involves 12\,$\mu$m data is roughly a factor 1.7 higher than the measured beam CO luminosities. This ratio is consistent with the $L_{\rm CO}^{\rm G}/L_{\rm CO}^{\rm B}$ ratio measured from ACA data, while the average ratio from CARMA galaxies is slightly higher, $\sim2$.}
    \label{F:apercorr}
\end{figure*}

\subsection{CALIFA IFU optical data}\label{SS:califa}
CALIFA\footnote{\url{https://califa.caha.es/}} is an IFU optical survey that observed hundreds of galaxies (considering both data release 3 and extended sample) using the PMAS/PPak integral field unit instrument of the 3.5m telescope hosted by the Calar Alto Observatory \citep{sanchez2012,sanchez2016b,lacerda2020,sanchez2023}. The CALIFA sample followed up Sloan Digital Sky Survey (SDSS, \citealt{york2000}) galaxies in order to be representative of the present-day galaxy population (0.001$<z<$0.08) in a statistically meaningful fashion (log($M_*/$[M$_{\odot}$])=9.4-11.4; E to Sd morphologies, with also irregulars, interacting, and mergers; \citealt{walcher2014,barrera-ballesteros2015}). Our sample considers the low-resolution (V500) set-up, which covers between 3745–7500$\,\AA$ with a spatial resolution FWHM=2.5 arcsec, and a spectral resolution  FWHM=6$\,\AA$. Note that recently, CALIFA has released an `extended and remastered' dataset called `eDR', which contains 895 collections of IFU maps having a spatial resolution FWHM$_{\rm PSF}\sim1$\,arcsec \citep{sanchez2023}. However, the \iedge\  is based on the extended CALIFA dataset first used in \cite{lacerda2020} and described further in \cite{villanueva2024}. This dataset has the same data quality as the CALIFA DR3 \citep{sanchez2016b}, though it includes more galaxies. CALIFA maps typically cover up to 2-2.5\,$R_{\rm eff}$ \citep{garcia-benito2015}.

\subsection{CARMA CO data}\label{SS:carma}
The first CO follow-up of CALIFA galaxies has been performed with the CARMA telescope and constitutes the original EDGE dataset of 178 galaxies mapped in $^{12}$CO(1-0) with the CARMA E-configuration (at a median sample resolution of 7.5 arcsec). A sub-sample of 126 of the higher signal-to-noise galaxies has also been observed in the D configuration, achieving a resolution of 4.5 arcsec (for the combined D+E dataset). CARMA was an interferometer constituted by 23 telescopes (with diameters of 3.5, 6.1, and 10.4\,m) located in the Owens Valley Radio Observatory (OVRO). Full details on the CARMA data have been presented in \cite{bolatto2017} (see also \citealt{wong2024} for a full description of the EDGE based on CARMA data\footnote{The current data release of the {\sc Python}-based EDGE is available at \url{https://github.com/tonywong94/edge_pydb}.}). As a reference, the average $\sigma_{\rm RMS}$ in the D+E galaxies is 38\,mK per 20\,km\,s$^{-1}$ channel width (or, given the average distances of the sample, a typical $1\sigma_{\rm RMS}$ surface molecular gas mass sensitivity $\Sigma_{\rm mol}=2.3\,$M$_{\odot}\,$pc$^{-2}$ for a 30\,km\,s$^{-1}$ line).

\subsection{ACA CO data}\label{SS:aca}
\cite{villanueva2024} presented a further follow-up of CALIFA data performed with ACA, dubbed ACA-EDGE. ACA is a sub-ALMA array composed of 12 7\,m antennas and 4 12\,m antennas organised on a compact configuration allowing the (partial) recovery of the extended emission. This sample consists of 60 galaxies mapped in $^{12}$CO(2-1) with $\sigma_{\rm RMS}\sim12-18$\,mK at 10\,km\,s$^{-1}$ channel width (e.g. in the range 0.9-1.2\,M$_{\odot}$\,pc$^{-2}$) and a spatial resolution between 5 and 7 arcseconds. The sample is drawn from galaxies with (10<log($M_*/$[M$_{\odot}$])<11.5 and declination $\delta<30^{\circ}$.

\subsection{APEX CO data}\label{SS:apex}
By number, the largest follow-up of CALIFA galaxies has been achieved with the APEX  12\,m telescope \citep{guesten2006}, where the $^{12}$CO(2-1) emission in 501 CALIFA galaxy central regions has been observed with a resolution of 26.3 arcsec. APEX is a single-dish telescope (basically a modified ALMA antenna) located at the Llano de Chajnantor Observatory in the Atacama Desert. 

This survey (dubbed APEX-EDGE) is fundamentally unbiased, as it targets objects constrained solely by the declination achievable by APEX (e.g. $\delta<30^{\circ}$). The APEX observations are single pointings that, given the APEX beam at 230\,GHz of 26.3 arcsec, cover approximately the central $1\,R_{\rm e}$ of the galaxies. In detail, the median $R_{\theta}/R_{\rm eff}=1.09$ with an interquartile range equal to 0.43.

The first 296 galaxies of this sub-sample were presented in \citet[][hereafter \citetalias{colombo2020}]{colombo2020}, as part of the M9518A\_103 and M9504A\_104 projects, while another sub-sample of 206 galaxies is being added here with data from the M9509C\_105, M9516C\_107, M9513C\_109 projects\footnote{ESO Program IDs: 0103.F-9518(A), 0104.F-9504(A), 0105.F-9509(C), 0107.F-9516(C), 0109.F-9513(C).} (PI: D. Colombo), which have been observed during the APEX 2020, 2021 and 2022 runs. The full project required approximately 580 hours of telescope time, including calibrations, overheads, and additional tests and off-centre observations (not presented here). The survey is designed to achieve a $\sigma_{\rm RMS}=2$\,mK (70\,mJy) at 30\,km\,s$^{-1}$ channel width. In this aspect, this survey is supposed to be CO flux-limited. However, for several observations, we have integrated longer to achieve better signal-over-noise (S/N) ratios. Given this, the median $\sigma_{\rm RMS}=1$\,mK with an interquartile range of 0.5\,mK (at 30\,km\,s$^{-1}$), corresponding to $M_{\rm mol}\sim0.4\times10^8$\,M$_{\odot}$ (or $\Sigma_{\rm mol}\sim0.6$\,M$_{\odot}$\,pc$^{-2}$) at 3$\sigma_{\rm RMS}$, assuming a constant $\alpha_{\rm CO(2-1)}=\alpha_{\rm CO(1-0)}/R_{21}=6.23$\,M$_{\odot}$\,(K\,km\,s$^{-1}$\,pc$^2$)$^{-1}$. With these criteria, we detected 321 (235) galaxies at S/N$>3$ (S/N$>5$), corresponding to 64\% (47\%) of the sample. 

M9518A\_130 and M9504A\_104 project observations were performed using PI230 single-pixel frontend. PI230 operates at a frequency of 230 GHz and employs dual-polarisation heterodyne technology. It features a bandwidth of 32 GHz and offers 524,288 spectral channels. The receiver spans a spectral range from  195 to 270 GHz and utilises a sideband-separating (2SB) mixer configuration.

M9509C\_105, M9516C\_107, M9513C\_109 project observations were acquired using nFLASH. While nFLASH features two independent frequency channels (nFLASH230 and nFLASH460), we have been using the 230\,GHz channel, nFLASH230, which is comparable to PI230. It operates at a spectral range from 196 to 281 GHz, covering 32 GHz IF instantaneous bandwidth including both sidebands and polarisations. The receiver shows a typical system temperature of 80-90 K.

To produce stable spectral baselines, the observations were performed using a symmetric wobbler-switching mode with a $100''$ chopping amplitude and a chopping rate of 1.5 Hz.
For each target, we used the coordinates provided for CALIFA galaxies and measured using Panoramic Survey Telescope and Rapid Response System (Pan-STARRS; \citealt{flewelling2020}) data (see \citealt{sanchez2023} for further details). The telescope was tuned to the COISO frequency (rest frequency 230.8 GHz in the USB) computed by using the stellar redshift inferred from CALIFA. For the survey, we adopted the typical APEX scheme recommended for ON-OFF observations. First, we start by adjusting the telescope focus in the $z$, $y$, and $x$ directions, with a preference for focusing on a planet; then we aligned the telescope pointing by referencing a bright source in close proximity to the target or, at the very least, locating one with a similar elevation in the sky. After these two steps, we performed the observing loop. The loop starts with a minute-long calibration scan (with the sequence on the `hot', `cold', and `sky' calibrators) and a 20-second-long ON-OFF scan repeated between 10-18 times, depending on the precipitable water vapour (PWV) level. The full loop (considering PWV) lasted between 14 to 25 minutes per science target. After typically 3-4 science observation loops ($\sim1$\,hour), we updated the pointing, while the focus was performed every 2 hours depending on changes in atmospheric conditions.

The APEX data calibration and reduction was performed using the Grenoble Image and Line Data Analysis Software (GILDAS\footnote{\url{http://www.iram.fr/IRAMFR/GILDAS}}) and the Continuum and Line Analysis Single-dish Software (CLASS\footnote{\url{https://www.iram.fr/IRAMFR/GILDAS/doc/html/class-html/class.html}}), wrapped into a {\tt python} routine, which employed pyGILDAS.\footnote{\url{https://www.iram.fr/IRAMFR/GILDAS/doc/html/gildas-python-html/gildas-python.html}} We used these tools to fit and eliminate a linear baseline from each spectrum located outside a 600\,km\,s$^{-1}$ window centred on the galaxy $V_{\rm LSR}$. The {\tt Python} wrapper was generally used for bookkeeping, for example to produce the final spectral table using astropy \citep{astropy2022}.\footnote{\url{https://docs.astropy.org/en/stable/api/astropy.table.Table.html}}

\section{Method}\label{S:methods}

\subsection{Database homogenisation}\label{SS:homogenisation}
As illustrated above, the data contributing to the \iedge\ are heterogeneous and require careful homogenisation to produce a unified database.
On the one hand, CARMA and ACA data are interferometric, where the CO emission has been mapped across the extent of the galaxies. On the other hand, APEX data are single-pointing observations, which collected the CO emission spectra from the central $\sim 1 R_\mathrm{eff}$ of the targets. Additionally, CALIFA data are also provided as pseudo-datacubes that include maps of various emission lines and stellar continuum quantities generated through the PIPE3D algorithm \citep[][]{sanchez2016b,sanchez2016c}. The homogenisation performed here implies a series of techniques. The \iedge\ includes two types of data: `beam' (or B) quantities, measured within the APEX beam of 26.3 arcseconds (e.g. within, approximately, the inner 1\,$R_{\rm eff}$ of the galaxies), and `global' (or G) quantities, integrated across the full galaxy extent (or within 2\,$R_{\rm eff}$, giving the typical footprint of CALIFA data). A similar dataset homogenisation has been used in \citetalias{colombo2020} and in the meantime, adopted also elsewhere (see \citealt{wylezalek2022}).

\subsubsection{Homogenisation of CALIFA data: Tapering}\label{SSS:califa_homo}
In order to infer beam quantities from CALIFA data, we used the same `tapering' technique defined in \citetalias{colombo2020}. We defined a `tapering' function $W_{\rm T}$, i.e. a bi-dimensional Gaussian,\footnote{Using {\sc astropy Gaussian2D} function, see \url{https://docs.astropy.org/en/stable/api/astropy.modeling.functional_models.Gaussian2D.html}} centred on the centre of the galaxy, with unit amplitude and FWHM $\theta$= $\sqrt{\theta_{\rm APEX}^2-\theta_{\rm CALIFA}^2}$, where the APEX beam FWHM $\theta_{\rm APEX}=26.3\,$arcsec, while $\theta_{\rm CALIFA}=2.5\,$arcsec. Given this, beam quantities are derived by spatially summing up all the pixels within the CALIFA maps (constructed through PIPE3D), previously multiplied by the Gaussian filter, $W_{\rm T}$. Averaged quantities within the APEX beam aperture are obtained using the weighted median of pixel values in the resolved maps where the weights for each pixel are given by the tapering function, $W_{\rm T}$.  This operation approximates the convolution of the CALIFA property maps to the APEX beam size and samples the result at the pointing centre of the APEX beam. 

Global quantities are calculated by integrating across the full CALIFA maps (see Section \ref{SS:quantities} for further details).

\subsubsection{Homogenisation of CARMA and ACA data: Smoothing}\label{SSS:carma_aca_homo}
To obtain beam quantities from CARMA and ACA datacubes, we adopted the same technique used recently by \cite{villanueva2024}. To do so, we followed the `smoothing' procedure from {\sc spectral-cube}.\footnote{\url{https://spectral-cube.readthedocs.io/en/latest/smoothing.html}} In essence, we convolved each channel map of the datacubes (using {\sc radio\_beam}\footnote{\url{https://radio-beam.readthedocs.io/en/latest/}}) with a bi-dimensional Gaussian with the size chosen to smooth the native beam sizes of the CARMA and ACA data to the size of the APEX beam at 230 GHz. Furthermore, we spectrally smoothed the derived datacubes to the spectral resolution of APEX data (30\,km\,s$^{-1}$) using the {\sc spectral\_smooth} functionality of {\sc spectral\_cube}. In this analysis, we used the unmasked datacube with a channel width of 20\,km\,s$^{-1}$. Afterwards, we extracted the central spectra from the spatially and spectrally smoothed cubes and we treated them similarly to APEX spectra (see Section~\ref{SS:quantities}). 

For the global quantities, we relied on the fluxes obtained using the smooth masking technique applied on both CARMA and ACA data (see \citealt{bolatto2017} for further details).

\subsubsection{Homogenisation of APEX data: Aperture correction}\label{SSS:apex_homo}
Beam quantities for APEX data are given by construction by the observations themselves.
In order to infer the global quantities, we have to impose an `aperture correction'. Aperture correction is a common practice for single-pointing data that (in most cases) do not completely cover the galaxies. Different studies have used several techniques to compensate for this (see e.g. \citealt{saintonge2011,bothwell2014}). Recently, \cite{leroy2021} analysed several of these aperture correction techniques to correct for the CO flux filtered out by PHANGS observations. They concluded that the correction produced through 12\,$\mu$m data is the most reliable as the correlation between CO intensity and the 12\,$\mu$m flux is the strongest analysed (see also \citealt{chown2021}). Given this, we downloaded z0MGS (z=0 Multiwavelength Galaxy Synthesis; \citealt{leroy2019}) data for 517 targets in our sample using the dedicated archive.\footnote{\url{https://irsa.ipac.caltech.edu/data/WISE/z0MGS/overview.html}} z0MGS is a collection of more than 15,000 galaxies observed in infrared and ultraviolet by the Wide-field Infrared Survey Explorer (WISE) and Galaxy Evolution Explorer (GALEX) missions. The 12\,$\mu$m emission data are extracted from the WISE W3 observations.

Afterwards, we followed several steps to generate the aperture correction for the APEX data. Firstly, we analysed the relationship between beam and global luminosities (see also Section~\ref{SS:quantities}) for CARMA, ACA, and WISE W3 data. To obtain beam fluxes for WISE W3 maps, we used the `tapering' approach adopted for CALIFA maps, considering the 7.5 arcseconds resolution data, which allows better separation between galaxy and foreground/background flux compared to the 15 arcseconds resolution data. Global fluxes from WISE W3 were calculated by co-adding all pixels within the central connected structure of the maps, masking all data below 3$\sigma_{\rm RMS}$, where $\sigma_{\rm RMS}$ is given by the `RMS' entry in the z0MGS-related data header. Foreground stars within z0MGS WISE W3 data were also masked out using the provided mask. From the 517 matched targets, we excluded galaxies that showed bad data quality after masking. Given this, we ended up with 494 WISE W3 targets to perform the aperture correction. We apply aperture corrections to the APEX data using this method, rather than directly relying on ACA or CARMA measurements, as the latter primarily sample star-forming, late-type galaxies, introducing a bias towards high specific star formation rates, for example. In contrast, the WISE photometry provides a more representative coverage of the parameter space sampled by the APEX galaxies, including objects in advanced stages of quenching. Nevertheless, we employ the CARMA and ACA datasets as benchmarks to evaluate the accuracy and consistency of the applied corrections, as detailed in the remainder of this section and in Section~\ref{SS:tel_comp}.

As shown in Fig.~\ref{F:apercorr} (left panel), the relationships between global and beam luminosities across the 3 datasets are broadly consistent and are in good statistical agreement. However, while the WISE W3 relation is linear, the relations for CARMA and ACA galaxies appear slightly sub-linear, possibly due to a lack of galaxies with low beam luminosity. The most significant deviations from the relationship are given by three galaxies in the CARMA sample (NGC5406, NGC1167, and NGC2916) that show global-over-beam luminosity ratios between $\sim11-25$. In those galaxies, indeed, the CO emission appears distributed in a ring, while the molecular gas is depleted in their centre (see \citealt{bolatto2017}, their Fig.~19-25). The relationships are measured from the detected 126 D+E CARMA, 60 ACA, and 494 WISE W3 galaxies. Data fitting is performed using \emph{linmix},\footnote{\url{https://linmix.readthedocs.io/en/latest/index.html}} which implies a hierarchical Bayesian model (by \citealt{kelly2007}) to perform a linear fit that takes into account uncertainties on both variables and upper limits (on the dependent measurements). Fundamentally, we obtained

\begin{equation}\label{E:wise_glob_beam_lum}
\log(L^{\rm WISE,~G}_{12\,\mu m})=0.05^{+0.05}_{-0.05}+1.02^{+0.01}_{-0.01}\log(L^{\rm WISE,~B}_{12\,\mu m}),
\end{equation}

\begin{equation}\label{E:aca_glob_beam_lum}
\log(L^{\rm ACA,~G}_{\rm CO(2-1)})=1.41^{+0.23}_{-0.23}+0.86^{+0.03}_{-0.03}\log(L^{\rm ACA,~B}_{\rm CO(2-1)}),
\end{equation}

\begin{equation}\label{E:carma_glob_beam_lum}
\log(L^{\rm CARMA,~G}_{\rm CO(1-0)})=1.44^{+0.18}_{-0.18}+0.87^{+0.02}_{-0.02}\log(L^{\rm CARMA,~B}_{\rm CO(1-0)}),
\end{equation}

where slopes and intercepts are given as medians of the respective posterior distributions, while the numbers in superscript (and subscript) represent the 75${\rm th}$ percentile minus the median (and the median minus the 25${\rm th}$ percentile).

The histograms in Fig.~\ref{F:apercorr} (lower left) show the ratio of the quantities in the upper panels. The ratios between global and beam quantities from WISE, CARMA, and ACA data can extend up to a factor of 30, but on average, they are distributed across factors $\sim1.6-2$. The median ratios for ACA and WISE data are identical. 

Furthermore, we measured the relationship between beam luminosities from APEX and WISE W3 data. This relationship from different S/N cuts of the APEX data is illustrated in Fig.~\ref{F:apercorr} (middle panel). In essence, we obtained:

\begin{eqnarray*}
\log(L^{\rm APEX\,(full),~B}_{\rm CO(2-1)})&=&-2.05^{+0.13}_{-0.14}+1.19^{+0.02}_{-0.02}\log(L^{\rm WISE,~B}_{12\,\mu m}),\\
\log(L^{\rm APEX\,(S/N>3),~B}_{\rm CO(2-1)})&=&-1.27^{+0.13}_{-0.13}+1.10^{+0.02}_{-0.01}\log(L^{\rm WISE,~B}_{12\,\mu m}),\\
\log(L^{\rm APEX\,(S/N>5),~B}_{\rm CO(2-1)})&=&-1.24^{+0.14}_{-0.14}+1.10^{+0.02}_{-0.02}\log(L^{\rm WISE,~B}_{12\,\mu m}).\\
\end{eqnarray*}

In particular, the last two relations are statistically consistent with the one derived by \citet{gao2019}, $\log(L_{\rm CO(2-1)}) = (-1.52 \pm 0.33) + (1.11 \pm 0.40)\log(L_{\rm 12,\mu m})$, based on a sample of 118 nearby galaxies with CO(2-1) detections — a sample approximately one-third the size of our APEX S/N>3 sub-sample, and about half the size of our S/N>5 sub-sample. It is also important to note that, unlike our analysis, \citet{gao2019} did not apply any tapering to the 12\,$\mu$m data, effectively comparing global $L_{\rm 12,\mu m}$ to central $L_{\rm CO(2-1)}$ measurements. 

The histograms in Fig.~\ref{F:apercorr} (lower middle) show that, despite considering APEX data sub-samples at different S/N thresholds (namely the full APEX data, or only detected galaxies with a S/N>3 and with S/N>5), the ratio between the APEX-measured beam CO(2-1) luminosity and the WISE-measured beam 12\,$\mu$m luminosity is remarkably similar with a typical value of $\sim0.45$. Finally, by combining the beam APEX CO(2-1)-WISE W3 12\,$\mu$m relationship, and the global-to-beam WISE W3 12\,$\mu$m relationship, we aperture-corrected the beam APEX CO(2-1) luminosities to infer the global APEX CO(2-1) luminosities. This gives

\begin{equation}\label{E:aper_corr}
\log(L^{\rm APEX, G}_{\rm CO(2-1)})=0.08+1.02\log(L^{\rm APEX, B}_{\rm CO(2-1)}).
\end{equation}

This relationship is largely comparable with the global-to-beam luminosity relationships inferred from CARMA and ACA data and shown in equations~\ref{E:carma_glob_beam_lum}-\ref{E:aca_glob_beam_lum}, and it is statistically consistent with the WISE global-to-beam luminosity relation (equation~\ref{E:wise_glob_beam_lum}).

The lower right panel of Fig.~\ref{F:apercorr} indicates that the global CO luminosities extrapolated for APEX data are a factor 1.7 larger than the measured APEX beam CO luminosities. This value is lower than the median from CARMA; however, it is identical to the median ratio inferred from ACA galaxies.

\subsection{Physical quantity derivation}\label{SS:quantities}
The physical quantities defined and used in this paper are substantially consistent with \citetalias{colombo2020}, with few modifications.

\subsubsection{CO observation-based quantities}\label{SSS:co_quantities}
For beam quantities from APEX, ACA, and CARMA observations, CO fluxes are derived within a spectral window of 600\,km\,s$^{-1}$ centred on the systemic velocity of the galaxy, which is measured from the stellar redshift. The CO line velocity-integrated flux density is expressed by the following equation:

\begin{equation}\label{E:fco}
    S_{\rm CO}\,\mathrm{[Jy\,km\,s^{-1}]} = \sum_i T^*_{{\rm A},i}\delta v .
\end{equation}

Here $\delta v=30\,$\kms\ is the channel width. While $T^*_{\rm A}$ is typically measured in K, we converted $S_{\rm CO}$ to Jy\,km\,s$^{-1}$ through a conversion factor between Kelvin and Jansky directly measured by APEX for each observing year using several calibrators.\footnote{\url{http://www.apex-telescope.org/telescope/efficiency/}} Since some of our observations span more than a single run, we considered an average of the fiducial conversion factors listed on the APEX website. We used a similar procedure to infer the proper main beam efficiency ($\eta_{\rm mb}$) for a given measurement set, in order to covert $T_{\rm mb}=T_{A}^*/\eta_{\rm mb}$. The median Jy/K conversion estimated in this way is 36, with an interquartile range equal to 2. To calculate the conversion factor for the interferometers, we simply assumed the beam and frequency information provided in the datacube headers. For ACA beam quantities, this factor is equal to 30.1, while for CARMA beam quantities, Jy/K=7.5. Note, however, that those conversion factors are related to the main beam temperature ($T_{\rm mb}$) scale, upon which the CARMA and ACA datacubes are provided. 

The statistical error for the flux can be measured as

\begin{equation}\label{E:eps_co}
    \epsilon_{\rm CO}\,\mathrm{[Jy\,km\,s^{-1}]} = \sigma_{\rm RMS}\sqrt{W_{50} \delta v}, 
\end{equation}

where $\sigma_{\rm RMS}$ is the standard deviation of the $T_{\rm A}^*$ variations measured in the first and last 20 line-free channels of each spectrum. As in \citetalias{colombo2020}, we used the second moment ($\sigma_{\rm v}$) calculated in the spectral window selected to measure the emission line to infer $W_{50} = \sqrt{8\log(2)}\sigma_{\rm v}$. Additionally, we assumed a fiducial $W_{50}=300$\,km\,s$^{-1}$, which is consistent with the peak of the $W_{50}$ distributions calculated for APEX, ACA, and CARMA-detected galaxies. The parameter $\epsilon_{\rm CO}$ is used to provide an upper limit of the flux for non-detected galaxies, where S/N = $T_{\rm A,~peak}^*/\sigma_{\rm RMS}<3$ (e.g. $S_{\rm CO}<3\epsilon_{\rm CO}$).

Given the flux, the CO luminosity can be inferred following equation 3 of \cite{solomon1997},

\begin{equation}\label{E:lco}
    L_{\rm CO}\,\mathrm{[K\,km\,s^{-1}\,pc^2]} = 3.25\times10^7\frac{D_L^2}{\nu_{\rm obs}^2 (1+z)^3}S_{\rm CO},
\end{equation}

where $D_L$ is the luminosity distance in Mpc (derived from the stellar redshift, $z$), $\nu_{\rm obs}$ is the observed frequency of the emission line in the rest frame in GHz ($\nu_{\rm obs}=115$\,GHz for CARMA data, and $\nu_{\rm obs}=230$\,GHz for APEX and ACA data), and $S_{\rm CO}$ is the CO velocity-integrated flux derived using equation~\ref{E:fco}.

This formula set is directly used to infer beam quantities. After that, in the case of APEX data, global $L_{\rm CO}$ are derived using the aperture correction given by equation~\ref{E:aper_corr}. Global fluxes are retroactively calculated from global luminosities. In this case, we assumed that the global and beam $\epsilon_{\rm CO}$ are equivalent. For CARMA and ACA data, we considered the published $S_{\rm CO}$ values to infer global luminosities (following equation~\ref{E:lco}). 

From the CO luminosity, the molecular gas mass, $M_{\rm mol}$, can be measured by assuming a CO-to-H$_2$ conversion factor, $\alpha_{\rm CO}$:

\begin{equation}\label{E:mmol}
    M_{\rm mol}\,\mathrm{[M}_{\odot}\mathrm{]} = \alpha_{\rm CO} L_{\rm CO}.
\end{equation}

More information regarding the $\alpha_{\rm CO}$ model adopted here is given in the next section.

\begin{figure}
    \centering
    \includegraphics[width = 0.4\paperwidth, keepaspectratio]{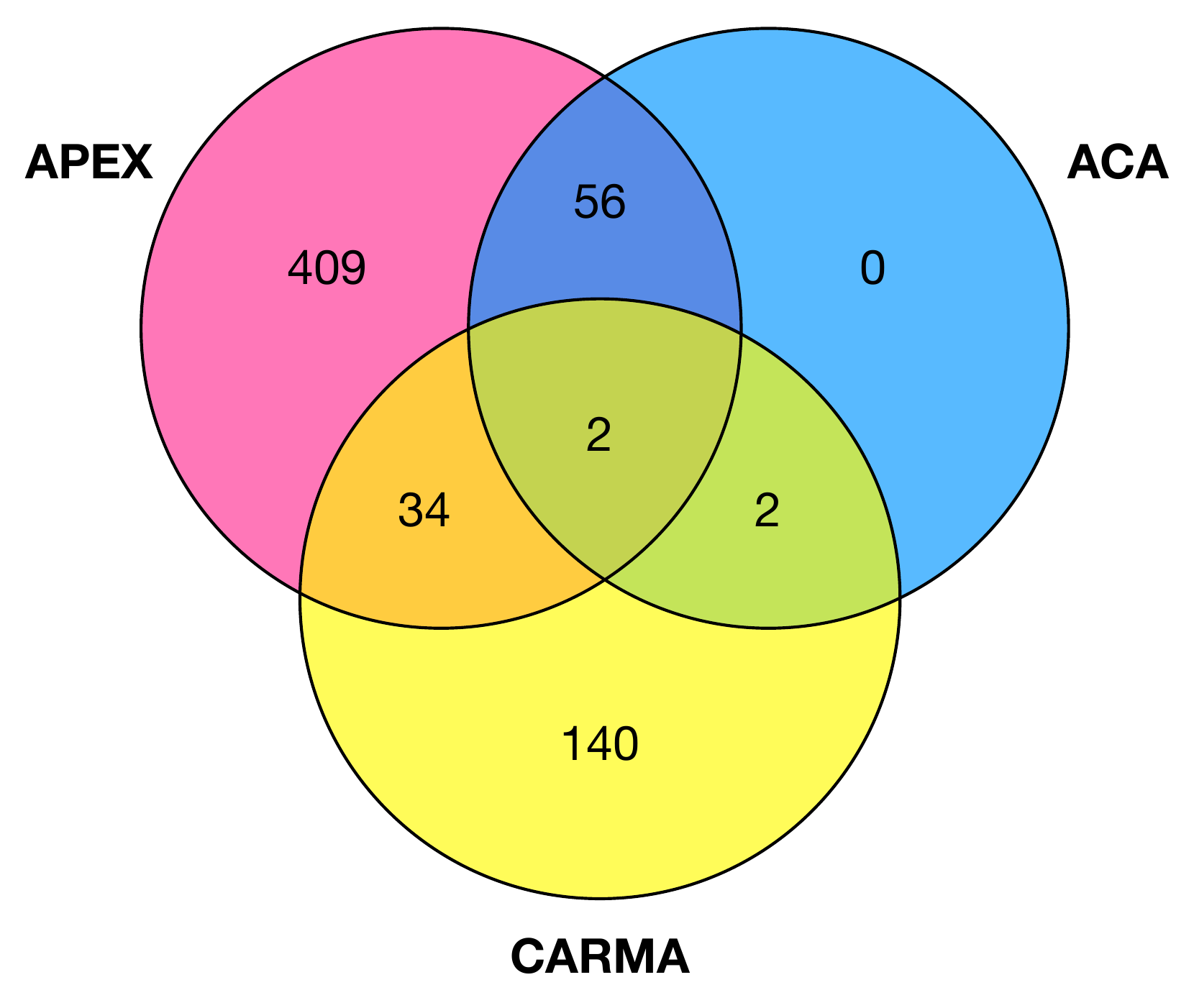}
    \caption{Number of galaxies within the \iedge\ observed by the different telescopes represented as a Venn diagram. Objects solely observed by APEX eventually constitute 64\% of the final database, and CARMA data add up to 22\% of the database. The rest of the targets have been observed by multiple telescopes.}
    \label{F:venn}
\end{figure}

\subsubsection{IFU observation-based quantities}\label{SSS:ifu_quantities}
As in \citetalias{colombo2020}, we used \ha, \hb, [OIII] $\lambda5007$, [NII] $\lambda6583$ flux maps, $F_{\rm H\alpha}$, $F_{\rm H\beta}$, $F_{\rm [OIII]}$, $F_{\rm [NII]}$, respectively; the \ha\ equivalent width, \wha, and the stellar mass surface density maps; all collected in pseudo-cubes, a term referring to the reduced data products produced by the PIPE3D pipeline \citep{sanchez2016b,sanchez2016c}. These pseudo-cubes are made of spatially resolved maps whose each spaxel contains physical quantities such as emission line fluxes, equivalent widths, and stellar mass surface densities derived from the analysis of CALIFA spectra. PIPE3D applies the GSD156 simple stellar populations (SSP) library \citep{cid_fernandes2013} to perform a spaxel-wise fit of the stellar population model on spatially rebinned V-band data cubes. This model is applied to calculate the stellar mass density value within each spaxel. The ionised gas data cube is inferred by subtracting the stellar population model from the original cube. Each of the 52 sets of emission line maps is obtained, calculating flux intensity, centroid velocity, velocity dispersion, and equivalent width for every spectrum. The maps included in the PIPE3D cubes are used to calculate integrated star formation rates (SFRs), stellar masses ($M_*$), and CO-to-H$2$ conversion factors ($\alpha_{\rm CO}$).

For the extinction-corrected SFR map derivations, we followed the procedure included in \cite{villanueva2024}, which is largely similar to \citetalias{colombo2020} but with a few modifications. In general, we applied the Balmer decrement method for each spaxel in the CALIFA map to calculate the nebular extinction

\begin{equation}\label{E:Aha}
    A_{\rm H\alpha}\,\mathrm{[Mag]} = \frac{K_{\rm H\alpha}}{0.4(K_{\rm H\beta} - K_{\rm H\alpha})}\times \log\left(\frac{F_{\rm H\alpha}}{2.86 F_{\rm H\beta}}\right),
\end{equation}

where the coefficients $K_{\rm H\alpha}=2.53$ and $K_{\rm H\beta}=3.61$ follow the \cite{cardelli1989} extinction curve \citep[see also ][]{catalan_torrecilla2015}.

The SFR follows as

\begin{equation}\label{E:sfr}
    \mathrm{SFR}\,\mathrm{[M_{\odot}\,yr^{-1}]} = 8\times 10^{-42} L_{\rm H\alpha}\times 10^{A_{\rm H\alpha}},
\end{equation}

as indicated by \cite{kennicutt1998}, which assumes the Salpeter initial mass function \citep{salpeter1955}, where $L_{\rm H\alpha}$ is the H$\alpha$ luminosity calculated as $L_{\rm H\alpha}$ [erg s$^{-1}$]=$4\pi D^2 F_{\rm H\alpha}$ [erg s$^{-1}$ cm$^{-2}$], with the distance to the target $D_{\rm L}$ in cm.

To obtain the final SFR maps, a series of masks is also considered. First of all, we masked all spaxels where S/N$<3$ for $H_{\alpha}$ and $H_{\beta}$. Additionally, we imposed a mask based on the $W_{\rm H_{\alpha}}$, where we considered only spaxels where $W_{\rm H_{\alpha}}\ge 6\,\AA$, as those are the only ones consistent with hydrogen ionisation from recent star formation activity (see \citealt{CidFernandes2010,sanchez2018,kalinova2021}). Furthermore, we applied a conservative mask that takes into account the spaxels where the ionisation is influenced by the AGN activity rather than purely by star formation. In essence, we removed all spaxels that lie above the BPT diagram \citep{baldwin1981} demarcation line given by \cite{kewley2001} in their equation 5. We also imposed several conditions to be fully consistent with \cite{villanueva2024} derivations. We considered a minimum $F_{\rm H\alpha}/F_{\rm H\beta}=2.86$, therefore we imposed $F_{\rm H\alpha}/F_{\rm H\beta}\equiv2.86$ for the spaxels where $F_{\rm H\alpha}/F_{\rm H\beta}<2.86$. For spaxels where $F_{\rm H\beta}$ is masked, therefore $A_{\rm H\alpha}$ not defined, we imposed $A_{\rm H\alpha}$ equal to the median of $A_{\rm H\alpha}$, calculated where $F_{\rm H\beta}$ is finite (as also assumed in \citealt{barrera-ballesteros2020}).

Finally, to obtain the integrated global SFRs, we spatially co-added the spaxels within the SFR maps, derived in equation~\ref{E:sfr} (namely SFR$_{\rm G}$=$\sum_i \mathrm{SFR}_i$, where $i$ runs across the spaxels of the map), while for the beam SFRs, before the summation, the maps are multiplied by the tapering function map, $W_{\rm T}$ (basically SFR$_{\rm B}$=$\sum_i \mathrm{SFR}_i\times W_{\rm T,i}$). 
\\

Similarly, the global stellar mass, $M_{\rm *,~G}$=$\sum_i$ $M_{{\rm *},i}$, is given by the summation over the whole map directly included in the PIPE3D pseudo-cube. The stellar mass within the APEX beam, $M_{\rm *,~B}$=$\sum_i$ $M_{{\rm *},i}\times W_{{\rm T},i}$, is obtained in the same way as SFR$_{\rm B}$ by the summation of the stellar masses from the spaxels, weighted by the tapering function $W_{\rm T}$. 
\\

Additionally, we used nebular emission lines information to infer the gas-phase metallicity, which is the fundamental parameter upon which $\alpha_{\rm CO}$ depends on. We measured the gas-phase metallicity using the O3N2 method by \cite{pettini_pagel2004}:

\begin{equation}\label{E:metallicity}
    12 + \log\left(\frac{\rm O}{\rm H}\right) = 8.73 - 0.32\times \log\left(\frac{F_{\rm [OIII]}}{F_{\rm H\beta}} \frac{F_{\rm H\alpha}}{F_{\rm [NII]}}\right).
\end{equation}

Nonetheless, \cite{pettini_pagel2004} O3N2 calibrator is obtained from HII regions. Therefore, we have to impose a cut on the metallicity maps based on \wha\ in order to exclude all non-star-forming regions (e.g. where $W_{\rm H\alpha}<6\,\AA$). To recover the metallicity in those regions, we created a map based on the metallicity-stellar mass relation (MZR). In particular, we used the model inferred from CALIFA data by \cite{sanchez2017}

\begin{equation}\label{E:mzr}
     12 + \log\left(\frac{\rm O}{\rm H}\right) = 8.73 + 0.01 \times (x-3.50)\exp(-(x-3.50)) ,
\end{equation}

where $x=\log(M_*/M_{\odot})-8.0$. This relation predicts the metallicity at 1\,$R_{\rm e}$ through the global stellar mass of the galaxy, $M_*$. We built a radial metallicity map assuming a universal metallicity gradient -0.1\,dex/$R_{\rm e}$ (see \citealt{sanchez2014} and \citealt{sun2020}). 

Several metallicity calibrations have been proposed (see \citealt{sanchez2019}, for example). Here, we adopt the \cite{pettini_pagel2004} calibration because it yields an oxygen abundance close to the Solar value assumed in this work \citep[8.69;][]{allende_prieto2001}. We emphasise, however, that the Solar metallicity is subject to uncertainties and revisions \citep[e.g.][]{lodders2003,asplund2009}, and our choice is intended to ensure a reasonable consistency with the assumed value. We explore the implications regarding the usage of these `combined' metallicity maps versus the `emission lines'-only derived metallicity maps on the derived \aco\ values in Appendix~\ref{A:aco}.

Beam median gas-phase metallicities are calculated assuming the weights given by the Gaussian filter, $W_{\rm T}$, on the combined metallicity map. Using this method, we obtain a median of 8.65 dex from the full galaxy sample with an inter-quartile range of 0.12 dex. The median beam metallicity with respect to the Solar metallicity is $Z'_{\rm B}=0.92$. Instead, the global metallicity is measured simply as the median across the full metallicity map. This gave us a median global metallicity across the full sample consistent with the beam metallicity, i.e. equal to 8.65 dex, with an interquartile range equal to 0.08. Similarly, we obtained $Z'_{\rm G}=0.80$.

As in \citetalias{colombo2020}, here we assumed a variable $\alpha_{\rm CO(1-0)}$ based on \cite{bolatto2013}, equation 31

\begin{equation}\label{E:aco}
    \alpha_{\rm CO(1-0)}\,\mathrm{[M_{\odot}\,(K\,km\,s^{-1}\,pc^{2})^{-1}]} = 2.9\exp\left(\frac{0.4}{Z'}\right)\times \left(\frac{\Sigma_{\rm *}}{100\,\mathrm{M_{\odot}\,pc^{-2}}}\right)^{-\gamma},
\end{equation}

\noindent where $Z'$ is the gas-phase metallicity relative to solar metallicity and $\Sigma_{\rm *}$ is the stellar mass surface density measured at each pixel in the CALIFA data and $\gamma=0.5$ where $\Sigma_{\rm *}>100$\,M$_{\odot}$\,pc$^{-2}$ or $\gamma=0$ otherwise. To avoid iterative solving, here we simply considered that $\Sigma_{\rm total}\equiv\Sigma_{\rm *}$, since for our sample of galaxies, the gas mass surface density is generally one order of magnitude lower than the stellar mass surface density. Also, $\Sigma_{\rm GMC}^{100}$, the Giant Molecular Cloud (GMC) molecular gas mass surface density in units of 100 M$_{\odot}$\,pc$^{-2}$,  does not appear in equation~\ref{E:aco} as we assume $\Sigma_{\rm GMC}^{100} = 1$ therefore, considering that GMC molecular gas mass surface density inner regions of nearby galaxies and Milky Way is largely consistent with 100\,M$_{\odot}$\,pc$^{-2}$ \citep[see ][]{sun2018,colombo2019,rosolowsky2021}. For low-mass galaxies (with $M_*<10^9$\,M$_{\odot}$) where the MZR is not defined and the optical emission lines remain undetected, we assumed $Z'=1$, which is the typical value for local Universe galaxies with Milky Way mass. In equation \ref{E:aco}, $\alpha_{\rm CO}$ is calculated across the whole CALIFA map; within the APEX beam aperture, we used the weighted median from the CO-to-H$_2$ conversion factor map, $\alpha_{\rm CO(1-0),~B}$, where the weights for each spaxel are given by the tapering function, $W_{\rm T}$. The global $\alpha_{\rm CO(1-0),~G}$ is simply the median across the $\alpha_{\rm CO(1-0)}$ map.
\\

The CO-to-H$_2$ conversion factor given by equation~\ref{E:aco} is directly applicable to calculate the molecular gas mass for CARMA observations, through equation~ \ref{E:mmol}. However, APEX and ACA observed the molecular gas through the 2-1 transition of $^{12}$CO. Generally, a constant CO(2-1)-over-CO(1-0) emission line ratio ($R_{21}\sim0.65$, \citealt{leroy2022}) is assumed to convert from the CO(1-0)-to-H$_2$ to the CO(2-1)-to-H$_2$ conversion factor (e.g. $\alpha_{\rm CO(2-1)} = \alpha_{\rm CO(1-0)} / R_{21}$). Nevertheless, den Brok et al. (in prep.) observed that, on kiloparsec scales, $R_{21}$ is tightly correlated to $\Sigma_{\rm SFR}$ inferred from $H\alpha$, and the relation appears to be environment-independent (see also \citealt{narayanan2014,leroy2022,den_brok2023}). In particular, they measured (their equation~4)

\begin{equation}\label{E:r21}
\log_{10}(R_{21})=0.12\log_{10}\left(\frac{\Sigma_{\rm SFR}}{\mathrm{M_{\odot}\,yr^{-1}\,kpc^{-2}}}\right)+0.06.
\end{equation}

We applied this relation to calculate $R_{21}$ maps from $\Sigma_{\rm SFR}$ maps. However, we used only spaxels where $\Sigma_{\rm SFR} > 10^{-3}$\, M$_{\odot}$\,yr$^{-1}$\,kpc$^{-2}$ to be consistent with the range of values considered by den Brok et al. (in prep.). As for $\alpha_{\rm CO}$, we assumed a global $R_{21,~G}$ as the median across the $R_{21}$ map, while $R_{21,~B}$ is again the median of $R_{21}$ map, where each spaxel is weighted by the tapering function $W_{\rm T}$. Given this, $\alpha_{\rm CO(2-1),~G} = \alpha_{\rm CO(1-0),~G} / R_{21,~G}$ and $\alpha_{\rm CO(2-1),~B} = \alpha_{\rm CO(1-0),~B} / R_{21,~B}$.
\\
Through these parameters, we had access to two fundamental quantities to study the star formation quenching across our galaxy sample, the star formation efficiency,

\begin{equation}\label{E:sfe}
\mathrm{SFE}=\frac{\mathrm{SFR}}{M_{\rm mol}},
\end{equation}

and the molecular gas mass-to-stellar mass ratio, or molecular gas fraction,

\begin{equation}\label{E:fmol}
f_{\rm mol}=\frac{M_{\rm mol}}{M_*}.
\end{equation}

\section{Results}\label{S:results}

\subsection{Comparisons between luminosities observed by different telescopes}\label{SS:tel_comp}

\begin{figure*}
    \centering
    \includegraphics[width = 0.85\paperwidth, keepaspectratio]{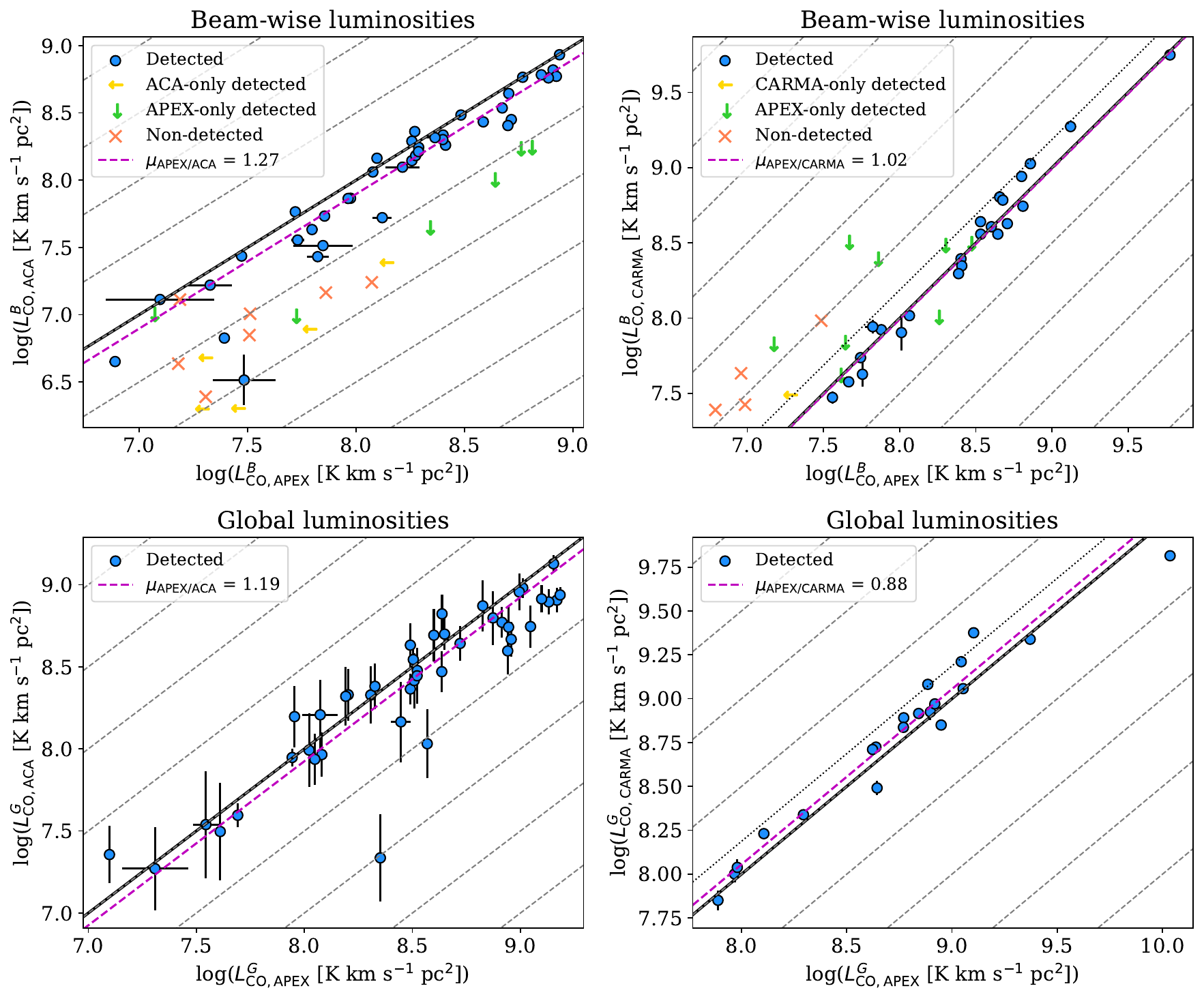}
    \caption{Comparison between beam-wise (top panels) and global luminosities (bottom panels) from ACA and APEX (left panels) and from CARMA and APEX (right panels). The blue circles indicate galaxies detected (with S/N > 3) from both telescopes, yellow arrows indicate galaxies detected only by ACA or CARMA, the green arrows indicate galaxies detected only by APEX, and the red crosses show galaxies not detected by both telescopes. For non-detections, luminosities are given as upper limits. For global luminosities, we show only the detections from both telescopes (see the text for further details). The black solid lines mark the 1:1 locus. The magenta dashed line has a unitary slope and an intercept equal to the median ratio between the x- and the y-quantities. In addition, in the left panels, the dotted black line marks the location where the CO(2-1)-to-CO(1-0) ratio, $R_{21}=0.65$. The grey dashed lines have a unitary slope and are separated by 0.5 dex.}
    \label{F:tel_comp}
\end{figure*}

Several targets in our sample have been observed by multiple telescopes, namely, 58 galaxies have been observed by both APEX and ACA, and 36 galaxies have been observed by both APEX and CARMA. This gives the opportunity to evaluate the global luminosities inferred for APEX galaxies and the performance of the aperture correction. Luminosities are calculated following the indications in Section~\ref{SS:quantities}.

The relations between beam luminosity inferred from APEX and ACA, and APEX and CARMA observations are shown in Fig.~\ref{F:tel_comp} (upper row). Generally, ACA beam luminosities follow APEX beam luminosities, with a smaller scatter compared to the other relations presented in this section. APEX luminosities are, on average, $\sim27\%$ higher than ACA luminosities. This can be due to differences in calibration, for example (as noticed by \citealt{villanueva2024}). Additionally, ACA data do not include total power observations, as the largest recoverable scale is basically the optical radius of the CALIFA galaxies (roughly 25 arcsec). However, we cannot rule out the possibility that a certain fraction of flux is filtered out in some galaxies. Larger deviations are observed for objects barely detected by ACA or highly inclined ones, for which the APEX beam extends further away from the galactic discs. Similarly, the luminosity measured for the same galaxies by CARMA and APEX is, on average, equal. However, while both ACA and APEX observed the CO(2-1) emission, CARMA observed the CO(1-0) emission. The typical CO(2-1)-to-CO(1-0) ratio, $R_{21}$, inferred from nearby spiral galaxies is $\sim0.65$ (see Section~\ref{SSS:ifu_quantities}). Here, instead, we measured a median $R_{21}\sim1$ in the centre of our galaxies. This higher-than-expected value can be attributed to differences in calibration between the two telescopes, however, the missing flux in the interferometric observations of CARMA might also play a role in increasing the ratio. As for the ACA sample, tests conducted to compare the CO fluxes between CARMA and COLDGASS\footnote{CO Legacy Database for GASS, \cite{saintonge2011}.} data did not reveal significant flux filtering, however, this cannot be completely excluded as CARMA data do not include a proper short-spacing correction (e.g. single-dish observations; see \citealt{bolatto2017}, their Section 3.2). Additionally, the three surveys have intrinsically been designed to reach different sensitivities. This is reflected in the upper limits of the luminosities. Indeed, considering the same galaxies, luminosity upper limits from APEX are more than 0.5 dex higher than the luminosity upper limits from ACA (Fig.~\ref{F:tel_comp}, upper left panel). On the opposite end, the luminosity upper limits from CARMA are 0.5 dex higher than the ones from APEX. 

The comparison between the global luminosities inferred from the three telescopes is shown in Fig.~\ref{F:tel_comp} (lower row panels). For this analysis, we do not consider the upper limits, as those are dependent upon the noise levels ($\sigma_{\rm RMS}$; equation~\ref{E:eps_co}), which in turn are dependent upon the (heterogeneous) channel widths of the original set of data. The global luminosities obtained through the aperture correction procedure described in Section~\ref{SSS:apex_homo} are largely similar to the global luminosities calculated from ACA data. Still, on average, APEX luminosities appear 16\% higher than ACA-inferred luminosities. Besides the effects that we described above regarding the beam-wise luminosity analysis, this discrepancy can also be due to the masking procedure adopted for ACA (and WISE W3) data that might provide slightly different fluxes based on the chosen noise threshold. However, it is important to note that the ratio between APEX and ACA global luminosities is roughly consistent with the same ratio measured for beam luminosities, meaning that the 12\,$\ mu$m-based aperture correction provides a robust recovery of the global flux for APEX observations. In essence, this test further validates our chosen aperture correction method for APEX galaxies. Instead, CARMA global luminosities appear, on average, 18\% higher than APEX global luminosities. Again, several effects could be involved in this discrepancy, together with the fact that the fluxes from the smooth masking procedure tend to be higher than the ones obtained from the more conservative dilated masking procedure (see \citealt{bolatto2017} for further details). In this aspect, however, from the global luminosities, we measure a median $R_{21}=0.82$, closer to the canonical $R_{21}=0.65$ value. Summarising, the comparison between observations from APEX, ACA, and CARMA demonstrates overall consistency in inferred CO luminosities, with systematic offsets possibly given by differences in calibration, sensitivity, and spatial filtering, and confirms the robustness of our aperture correction method for APEX data.

\subsection{Distance bias}\label{SS:dist_bias}
\begin{figure}
    \centering
    \includegraphics[width = 0.4\paperwidth, keepaspectratio]{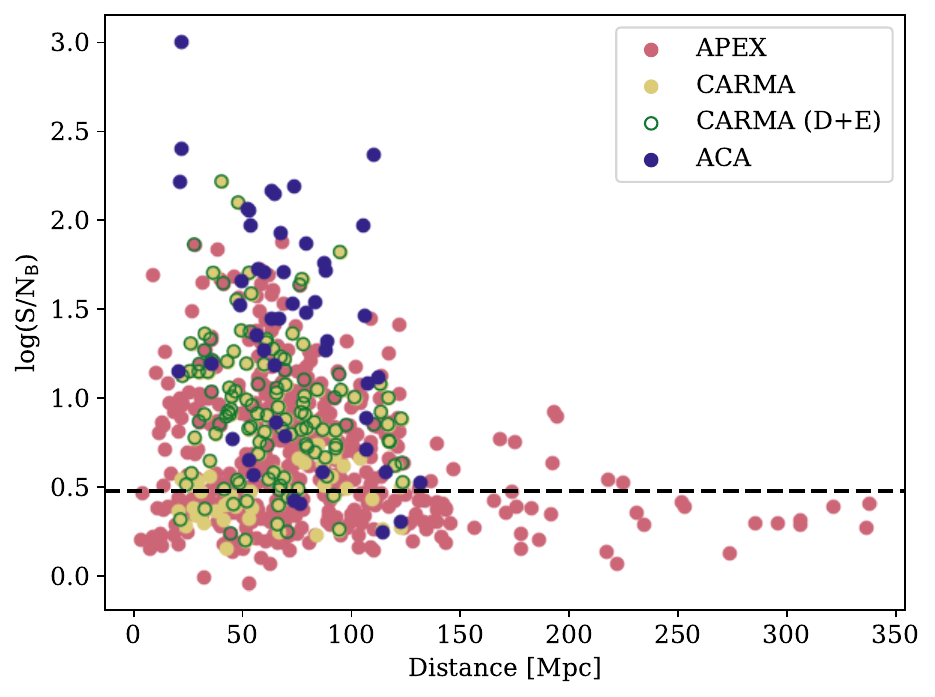}
    \caption{Detection distance bias shown as the beam S/N vs the distance to the galaxies, for APEX- (red), CARMA- (yellow), CARMA (D+E)- (green empty), ACA- (blue) observed galaxies. The black dashed line indicates a S/N=3 that marks the separation between detections and non-detections. Non-detections (observations with S/N<3) starkly decrease from a distance >130 Mpc and are all related to APEX observations.}
    \label{F:dist_bias}
\end{figure}

Although CALIFA galaxies are all local objects (having a maximum redshift equal to 0.08) and are diameter-selected, some distance bias can be present in our database. In Fig.~\ref{F:dist_bias}, the beam S/N ratio of the CO observation versus the distance to the galaxies is shown for the different telescopes constituting the final sample. The bulk of objects is observed within a distance of 150 Mpc. In this distance range, the number of the detected galaxies is a factor 2.4 larger than the number of non-detected galaxies. However, for distances >130 Mpc, the galaxies are mostly non-detected, and they are all observed by APEX. Nevertheless, those are only 50 targets or 8\% of the sample. This concludes that a significant distance detection bias is absent in our database.

\subsection{\iedge\ sample statistics}\label{SS:sample_stat}

\subsubsection{Definition of the consolidated sample}\label{SSS:sample_stat_unique}

\begin{figure*}
    \centering
    \includegraphics[width = 0.85\paperwidth, keepaspectratio]{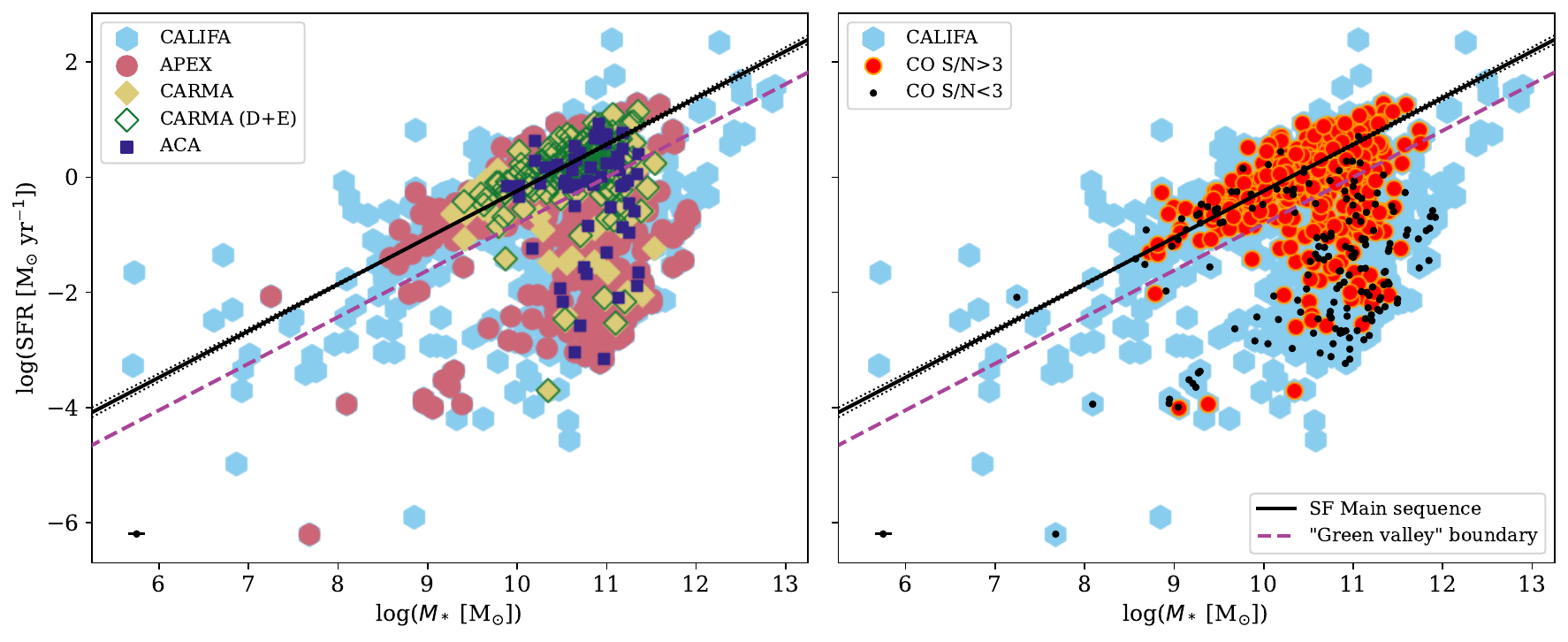}
    \caption{SFR-$M_*$ diagrams defined by the different samples considered in this paper (left panel): full CALIFA (cyan hexagons), and galaxies observed in CO lines by APEX (red circles), CARMA (yellow diamonds), CARMA (D+E galaxies only, empty green diamonds), ACA (blue squares). In the right panel, the red circles represent CO-detected galaxies (with S/N > 3), while the black dots show CO non-detections (with S/N < 3). In both panels, the black line shows \cite{cano_diaz2016} the SFMS model with the uncertainties (dotted lines). The dashed purple line indicates the green valley boundary position defined in \citetalias{colombo2020}, which is located $3\sigma$ (equal to 0.20 dex as in \citealt{cano_diaz2016}) from the SFMS. APEX targets extend from the SFMS to the retired regions, better representing the CALIFA galaxies in this diagram.}
    \label{F:sfm_sample_det}
\end{figure*}

The \iedge\ that we are presenting here consists of galaxies which have been observed by multiple telescopes. The Venn diagram represented in Fig.~\ref{F:venn} summarises the sample overlaps. The majority of targets ($\sim64\%$) in the database have been observed only by APEX, while another $\sim22\%$ has been uniquely observed by CARMA.  Almost all ACA galaxies (58 targets) were previously observed by APEX. Additionally, 34 targets have been observed by both APEX and CARMA, while 2 galaxies (UGC08781 and UGC10972) have been observed by all three telescopes. If we restrict the CARMA data to the D+E targets, the distribution of the target number changes slightly. CARMA D+E galaxies constitute $\sim16\%$ of the database, while the overlap with APEX is constituted by 26 galaxies. With ACA, we did not reobserve D+E targets (meaning there is no overlap between the D+E CARMA and ACA), but ACA observed 2 CARMA E-only galaxies (NGC2449, UGC05396), which have not been observed by APEX.
\\

The distribution of galaxies across the SFR-$M_*$ diagram is typically used as a standard diagnostic for sample characterisation; therefore, in Fig.~\ref{F:sfm_sample_det} (left), we present the position of the various galaxy samples within our database on this diagram. It is clear that the APEX sample covers the parameter space defined by the two variables, and it is more representative of the CALIFA sample compared to the others. In particular, while CARMA galaxies extend from the star formation main sequence region to the retired region, quenched galaxies are mostly observed by the E-configuration only, while D+E targets are constrained to the SFMS. ACA targets seem to follow a distribution similar to the CARMA E targets. The distribution of the targets on the SFR-$M_*$ diagram considering their S/N is illustrated in Fig.~\ref{F:sfm_sample_det} (right). It is interesting to notice that while the number of detected galaxies is generally found along the SFMS, several non-detections are also observed in the same region. Still, there are several detections in the green valley and the retired region of the diagram.\\

\begin{figure*}
    \centering
    \includegraphics[width = 0.85\paperwidth, keepaspectratio]{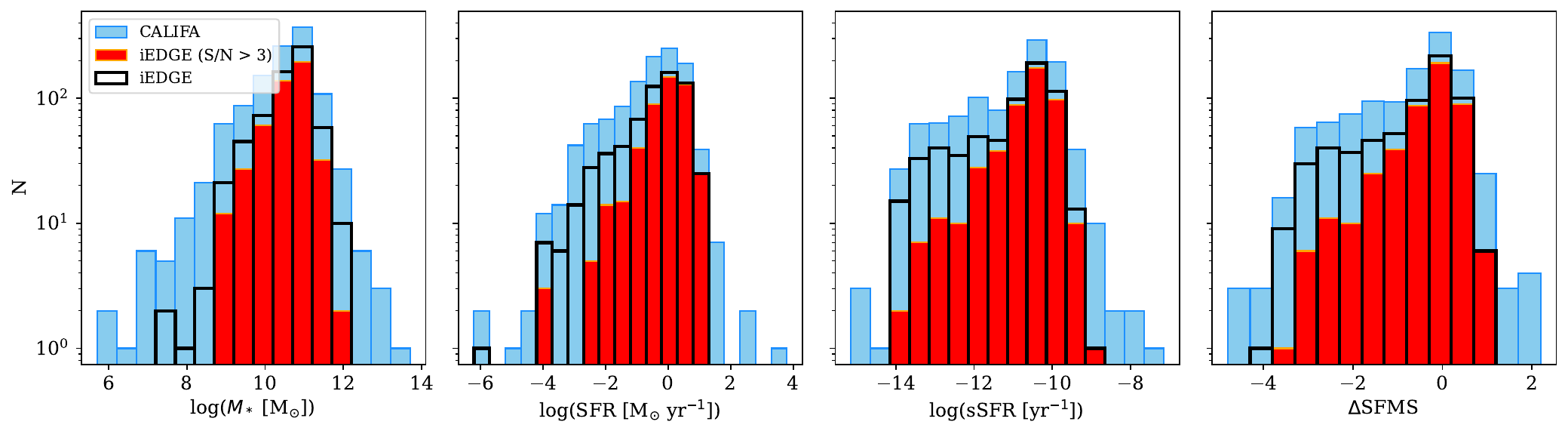}
    \caption{Histograms across samples (full CALIFA, cyan bars; galaxies in the full \iedge\ sample, black empty bars; CO detected galaxies in the \iedge, red bars) related (from left to right) to $M_*$, SFR, specific SFR (sSFR=SFR/$M_*$), and logarithmic distance from the SF main sequence ($\Delta$SFMS). Similarly to CALIFA, \iedge\ spans a representative range of SFR, $M_*$, and related values. However, CO detections decrease at low sSFR and $\Delta$SFMS values.}
    \label{F:sfr_mstar_histo}
\end{figure*}

Given the overlaps described at the beginning of this section, in the analyses adopted in the paper, we have used a set of criteria to define our `consolidated' sample of unique galaxies. For consistency, we prioritised APEX data and used CARMA data only if the targets had not been covered by APEX. We used ACA data instead of APEX data in case the latter showed better S/N compared to the former. We had, therefore, included 454 APEX galaxies, 47 ACA galaxies, 100 CARMA (D+E) galaxies, and 42 CARMA (E) galaxies. Nevertheless, we verified that for galaxies observed by multiple telescopes (64 objects), the mean ratio between these measurements and the corresponding values in the consolidated sample remains close to unity (on average). The most significant discrepancies (up to a factor of 2) occur when the S/N of the measurements contributing to the mean is either low (S/N$\sim$3–4, as in the case of NGC 0155) or when the individual measurements exhibit substantial differences due to the S/N of the detections (e.g. UGC 08781). Finally, the description of all quantities included in this release of the database is in Table~\ref{T:database_entry}.\\

\subsubsection{Star formation rate and stellar mass completeness}\label{SSS:sample_stat_sfr_mstar}

\begin{figure*}
    \centering
    \includegraphics[width = 0.85\paperwidth, keepaspectratio]{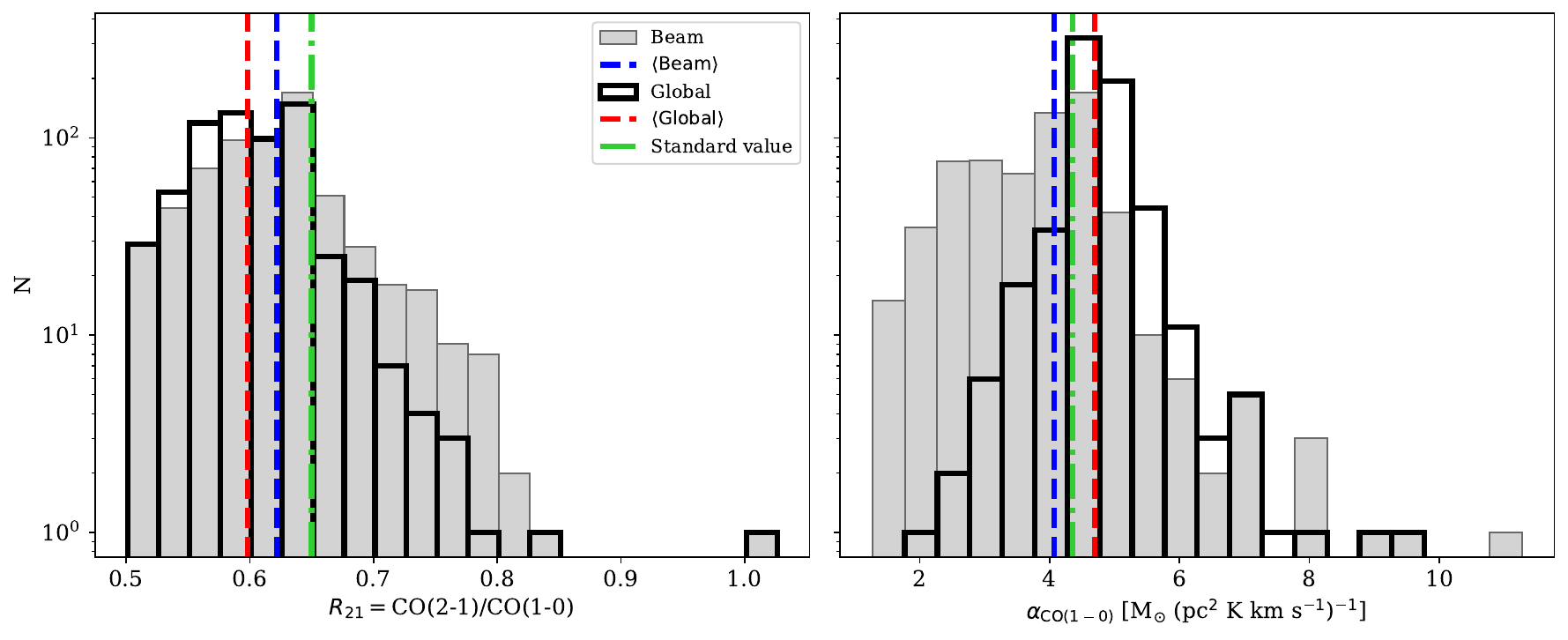}
    \caption{Distributions of the inferred CO(2-1)-to-CO(1-0) ratio ($R_{21}$, left) and CO(1-0)-to-H$_2$ conversion factors ($\alpha_{\rm CO(1-0)}$, right). These quantities are derived using IFU maps from models described in equations \ref{E:aco} and \ref{E:r21} (for $\alpha_{\rm CO(1-0)}$ and $R_{21}$, respectively). Histograms show median $R_{21}$ and $\alpha_{\rm CO(1-0)}$ calculated across the full maps (global, empty black histograms) and tapered maps (beam, grey histograms). The red (blue) dashed lines indicate the median of the global (beam) quantities, while the green dash-dotted lines show the typical values adopted for star-forming galaxies in the local Universe (e.g. $R_{21}=0.65$ and $\alpha_{\rm CO(1-0)}=4.35$ M$_{\odot}$\,(pc$^2$\,K\,km\,s$^{-1}$)$^{-1}$). Regarding $R_{21}$, we measured medians (depending on the coverage) generally lower than the average values observed in nearby galaxies. The $\alpha_{\rm CO(1-0)}$ values span an order of magnitude, with lower conversion factors measured within the galaxy centres. This is consistent with the fact that galaxies exhibit a declining metallicity trend with respect to the galactocentric radius, hence showing lower $\alpha_{\rm CO(1-0)}$ in their centre (see also Appendix~\ref{A:aco}).}
    \label{F:r21_aco_histo}
\end{figure*}

The statistics of SFR and $M_*$-related properties for our consolidated samples compared with the parental CALIFA sample are collected in Fig.~\ref{F:sfr_mstar_histo}. Our CO-detected targets showed a stellar mass, which generally missed the low and high mass tails of the CALIFA distribution, and they are generally limited to $8.5<\log(M_*/\mathrm{[M_{\odot}]})<12.5$. The full \iedge\ sample contains larger fractions of data below and above those limits, but they are not detected. Regarding the SFR, our detections are representative of the $-1<\log(\mathrm{SFR/[M_{\odot}\,yr^{-1}]})<1$  range of the CALIFA parental distribution. Several targets are observed below  $-1<\log(\mathrm{SFR/[M_{\odot}]}$, but they are not detected in CO. This might simply reflect the fact that these targets are quenched due to low molecular gas fractions. This evidence is reflected in the sSFR histogram (see Fig.~\ref{F:sfr_mstar_histo}). Although galaxies with low sSFR are included in our sample, the \iedge\ misses objects with $\log(\mathrm{sSFR/[yr^{-1}]})>-9$. Considering the logarithmic distance from the SFMS fit (for which we assumed the \citealt{cano_diaz2016} model: $\log(\mathrm{SFR/[M_{\odot}\,yr^{-1}]})=-8.34+0.81\log(M_*/\mathrm{[M_{\odot}]})$), it seems that these high sSFR galaxies are objects more than 1 dex away from the SFMS, or that the \iedge\ sample generally missed the very high star-forming galaxies (possibly starburst, where $\Delta\mathrm{SFMS}>1$) present in the CALIFA parental sample. Additionally, the database does not contain the few low ($\Delta\mathrm{SFMS}<-4$) targets present in the CALIFA sample. This said, the \iedge\ is largely representative of the CALIFA parental sample considering sSFR and $\Delta\mathrm{SFMS}$, which are fundamental to studying star formation quenching in the galaxies.

\subsubsection{$R_{21}$ and $\alpha_{\rm CO(1-0)}$ distributions}\label{SSS:sample_stat_r21_aco}
As described in Section~\ref{SS:quantities}, the IFU data provide a way to model several quantities that are generally assumed when multiple CO transition and isotopologue emissions are not available to properly infer the H$_2$ column density, namely $R_{21}$ and $\alpha_{\rm CO(1-0)}$. Here, for each galaxy in the sample, we assumed $R_{21}$, calculated through its correlation with the SFR surface density (see equation~\ref{E:r21}). The distribution of this (global and beam) parameter is shown in Fig.~\ref{F:r21_aco_histo} (left) and ranges between 0.5-0.85 (with a single object showing $R_{21}>1$). In detail, we obtained a median $R_{21,~G}=0.60$ with an interquartile range of 0.08 and a median  $R_{21,~B}=0.62$ with an interquartile range equal to 0.07. The medians, in particular, are consistent with the $R_{21}$ values measured across galactic discs via direct comparison of the CO(2-1) and CO(1-0) emission of 0.6-0.70 (see \citealt{yajima2021,den_brok2021,leroy2022}). In Section~\ref{SS:tel_comp}, we calculated a median $R_{21}=0.82$ for the galaxies for which we possess both APEX and CARMA data. However, several objects showed luminosities ratios close to the canonical $R_{21}=0.65$, with other galaxies deviating from this value. Nevertheless, $R_{21}\sim0.8$ are not uncommon in the literature, as reported by several studies \citep{braine1993,saintonge2017,cicone2017}.

The distribution of global and beam CO(1-0)-to-H$_2$ conversion factors ($\alpha_{\rm CO(1-0)}$) modelled from the gas-phase metallicity and the stellar mass surface density (following equation~\ref{E:aco}) is shown in Fig.~\ref{F:r21_aco_histo} (right). Together, histograms spread between $\sim1.5-11$. In detail, we obtained a median $\alpha_{\rm CO(1-0),~B}=4.07\,\mathrm{M_{\odot}\,(K\,km\,s^{-1}\,pc^{2})^{-1}}$, with an interquartile range equal to $1.46\,\mathrm{M_{\odot}\,(K\,km\,s^{-1}\,pc^{2})^{-1}}$; and a median $\alpha_{\rm CO(1-0),~G}=4.70\,\mathrm{M_{\odot}\,(K\,km\,s^{-1}\,pc^{2})^{-1}}$, with an interquartile range equal to $0.45\,\mathrm{M_{\odot}\,(K\,km\,s^{-1}\,pc^{2})^{-1}}$. Both estimations are statistically consistent with the Milky Way generally assumed value $\alpha_{\rm CO(1-0)}=4.35\,\mathrm{M_{\odot}\,(K\, km\,s^{-1}\,pc^{2})^{-1}}$. We discuss further implications on the assumed method to calculate \aco\ and the dependences of \aco\ on metallicity and stellar mass surface density in Appendix~\ref{A:aco}.

\subsection{Fundamental scaling relations for star formation}\label{SS:results_relations}
\begin{figure*}
    \centering
    \includegraphics[width = 0.9 \paperwidth, keepaspectratio]{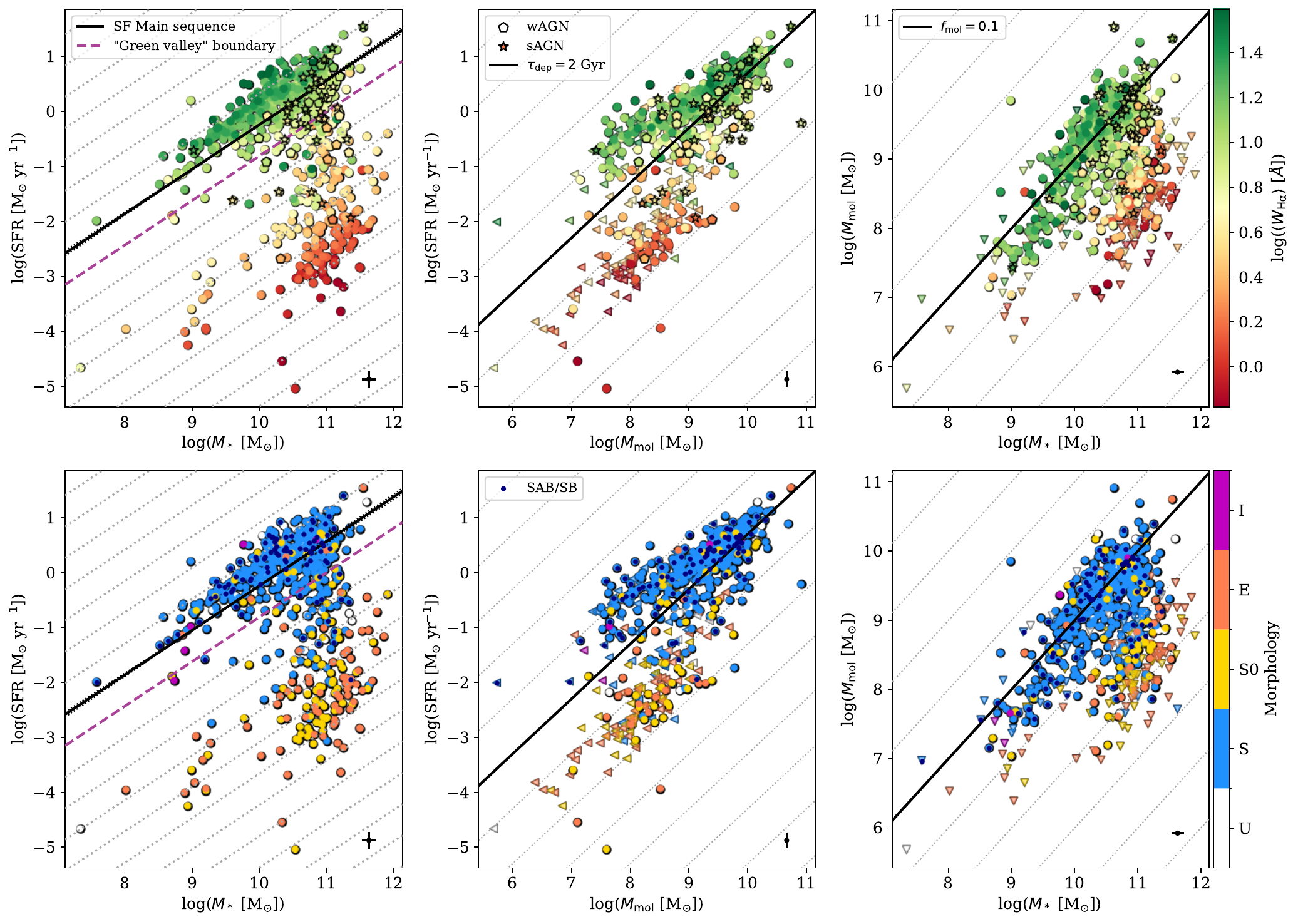}
    \caption{Star formation scaling relations studied here colour-coded by the median equivalent width of H$_{\alpha}$ ($W_{\rm H\alpha}$, upper row) and morphology (lower row), from left to right SFR-$M_*$, SFR-$M_{\rm mol}$, and $M_{\rm mol}-M_*$. In the left panel, the elements follow Fig.~\ref{F:sfm_sample_det}; additionally, the dotted lines are spaced of $3\sigma$ from each other. In the middle panel, the black solid line indicates a constant molecular depletion time $\tau_{\rm dep}=2$\, Gyr, additionally, dotted lines show constant $\tau_{\rm dep}$ from (top to bottom) 10$^{-5}$\, Gyr to 10$^7$\, Gyr. In the right panel, the black line indicates a constant molecular gas-to-stellar mass $f_{\rm mol}=0.1$; additionally, the dotted lines show constant $f_{\rm mol}$ from (bottom to top) 10$^{-5}$ to 10$^4$. The triangles in the middle and right panels illustrate galaxies for which upper limits for the molecular gas mass are used. At the bottom right of each panel the typical (median) uncertainty for the quantities is shown. In the upper row, the pentagons show the positions of the wAGN-hosts in the sample, while the stars show the positions of sAGN-hosts. In the lower row, the dark blue dots show the locations of the SAB/SB galaxies. Spiral star-forming galaxies on the one hand and early-type retired galaxies on the other hand describe well-defined relationships across the three diagrams.}
    \label{F:scalrels_wha_morph}
\end{figure*}

The scaling relations between SFR, $M_*$, and $M_{\rm mol}$ can provide valuable insight into the evolution of galaxies. With the 643 galaxies included in the \iedge\ database, we can accurately represent them (Fig.~\ref{F:scalrels_wha_morph}).
For this analysis, we colour-encoded the diagram based on the median H$\alpha$ equivalent width across the full maps, $\langle W_{\rm H\alpha}\rangle$ (Fig.~\ref{F:scalrels_wha_morph}, upper row) and on their morphology (Fig.~\ref{F:scalrels_wha_morph}, lower row). Generally, star-forming, spiral galaxies and retired, early-type galaxies form well-distinguishable sequences across the three diagrams. In the following, we also consider the nuclear activity of the galaxies. To establish if a galaxy hosts an AGN, we adopted the scheme by \cite{kalinova2021} (but see also \citealt{CidFernandes2010,sanchez2014,lacerda2018,lacerda2020}), who considered the three Baldwin-Phillips-Terlevich (BPT; \citealt{baldwin1981}) diagnostic diagrams that use the [OIII], [SII], [OI] and [NII] line ratios with respect to the H$\alpha$ and H$\beta$ lines. In this scheme, if three spaxels (covering a CALIFA PSF) within 0.5\,$R_{\rm e}$ are found in the Seyfert region of at least 2 BPT diagrams, the galaxy is considered a candidate AGN-host. Nevertheless, the spaxels need to show a $W_{\rm H\alpha}\ge3$\,\AA\ to be considered as a weak AGN (wAGN) host galaxy, or a $W_{\rm H\alpha}\ge6$\,\AA to be considered as a strong AGN (sAGN) host galaxy.

The standard appearance of the SFR$-M_*$ diagram \citep[e.g.][]{brinchmann2004,daddi2007,catalan_torrecilla2015,renzini_peng2015,cano_diaz2016,sanchez2018,sanchez2020} is faithfully reproduced by the \iedge\ data (Fig.~\ref{F:scalrels_wha_morph}, left), with galaxies across the SFMS having $\langle W_{\rm H\alpha}\rangle\ge6\,\AA$ in general, galaxies in the red sequence showing typically $\langle W_{\rm H\alpha}\rangle\leq3\,\AA$, and green valley targets bounded by these two limits. Still, we found several galaxies having large $\langle W_{\rm H\alpha}\rangle$ values up the retired region. As it has been established several times, $W_{\rm H\alpha}$, along side the colour, is a faithful proxy of the star-forming properties of galaxies. Indeed, finding $\langle W_{\rm H\alpha}\rangle\ge6\,\AA$ indicates hydrogen ionisation due to star formation activity (e.g. HII regions),  while $\langle W_{\rm H\alpha}\rangle\le3\,\AA$ indicates ionisation due to the old stellar population \citep[e.g.][]{sanchez2020, kalinova2021}. Therefore, it is not surprising to find that  spiral and star-forming galaxies, on the one hand, and retired spheroids, on the other hand, are characterised by those values of $W_{\rm H\alpha}$. Active galaxies mainly occupy the high mass region of the diagram (typically $M_*>10^{10}$\, M$_{\odot}$) and they are constrained between the SFMS fit and the red sequence (where SFR$<10^{-2}$\, M$_{\odot}$\,yr$^{-1}$). Across the diagram, spiral galaxies appear to dominate the SFMS and the green valley, while the retired region is constituted by early-type galaxies. Still, we observe that some S0-type objects have high SFR. Barred galaxies are mostly observed above the SFMS fit. 
In the SFR$-M_{\rm mol}$ relation (Fig.~\ref{F:scalrels_wha_morph}, middle) star-forming galaxies (with $\langle W_{\rm H\alpha}\rangle\ge6\,\AA$) are organised across a line defined by a molecular depletion time $\tau_{\rm dep}=2$\, Gyr, which is considered as the typical depletion time for star-forming galaxies \citep[e.g.][]{bigiel2011,rahman2012,leroy2013,utomo2017,sun2023}. However, we also noticed a deviation towards higher SFR (therefore lower $\tau_{\rm dep}$) for star-forming galaxies with $\log(M_{\rm mol}/\mathrm[M_{\odot}])\sim8$. This deviation might be driven by low-mass galaxies, while the most gas-rich objects appear well-organised across the 2\, Gyr line. The same deviation can be noticed in \cite{villanueva2021} (see their Fig.~7), who reproduced the same plot using ACA and CALIFA data. This region is clearly dominated by spiral galaxies (Fig.~\ref{F:scalrels_wha_morph}, middle bottom), although several S0 objects are observed across the region. This is not uncommon, though. Several early studies have found early-type galaxies with SFE comparable to spirals \citep[e.g.][]{lees1991,wiklind1995,combes2007}. The equivalent of the green valley in this diagram is not clearly noticeable, as galaxies with $3<\langle W_{\rm H\alpha}\rangle<6\,\AA$ look to be distributed among the star-forming ones. Instead, retired targets appear to be well separated from the rest, forming an independent `sequence' apparently characterised by $\tau_{\rm dep}\sim50-100$\, Gyr. These depletion times are consistent with the value measured for retired targets in the xCOLDGASS sample \citep{saintonge2017}. This sequence appears to be formed solely by early-type objects (E and S0). Nevertheless, it is difficult to provide a robust definition of this sequence as many targets are non-detected, and we showed only an upper limit for these targets in this diagram. Active galaxies are organised with generally high molecular gas mass and SFR ($M_{\rm mol}>10^9$\, M$_{\odot}$ and SFR$<10^{-1}$\, M$_{\odot}$\,yr$^{-1}$, respectively). Barred galaxies are located mainly above the line that defined the $\tau_{\rm dep}=2$\, Gyr. In the rightmost panels of Fig.~\ref{F:scalrels_wha_morph}, the logarithmic $M_{\rm mol}-M_*$ relationship is displayed. As in the previous diagrams, sequences for the star-forming and retired galaxies are well defined, while the transitional region (the equivalent of the `green valley' in the SFR$-M_*$ relation) is not clearly distinguishable. In particular, the sequence for galaxies dominated by star formation is characterised by $f_{\rm mol}\sim0.1$, even if galaxies above and below $\log(M_*/\mathrm{[M_{\odot}]}\sim10)$ show somehow $f_{\rm mol}$ below and above this value. Instead, the retired region $f_{\rm mol}$ is at least below 0.01. This result is consistent with the measurements from the ATLAS3D survey \citep{young2011} reanalysed by \cite{saintonge_catinella2022}.  As noticed before, the active galaxies are generally located at the high $M_{\rm mol}$ and $M_*$ ends of the diagram (see also \citealt{sanchez2018}). Spiral galaxies dominate the `molecular gas main sequence' of the diagram, while early-type objects are organised across the retired galaxy region.

Our results are qualitatively consistent with the kiloparsec-scale resolved analyses conducted by the EDGE-CALIFA and ALMaQUEST projects on the same topic. Both star-forming galaxies \citep{bolatto2017,utomo2017,lin2019,baker2022,villanueva2024} and individual star-forming spaxels \citep{ellison2021a,ellison2021b,lin2022} follow well-defined sequences in the resolved SFR–$M_{\rm mol}$ and $M_{\rm mol}$–$M_*$ diagrams. In contrast, spaxels in green valley and retired systems \citep{lin2022,villanueva2024}, as well as retired spaxels specifically \citep{ellison2021b,lin2022}, exhibit systematically lower SFE and \fmol\ compared to their star-forming counterparts. This confirms the conclusions of \citet{sanchez2020}, who demonstrated that global and kiloparsec-scale measurements follow statistically consistent scaling relations and extend them to the quiescent regime, where both global and local measurements appear to trace the same underlying relations.

\subsection{Global versus inner galaxy behaviours of star formation efficiency and molecular gas fraction}\label{SS:results_globbeam_sfe_fmol}
\begin{figure*}
    \centering
    \includegraphics[width = 0.85 \paperwidth, keepaspectratio]{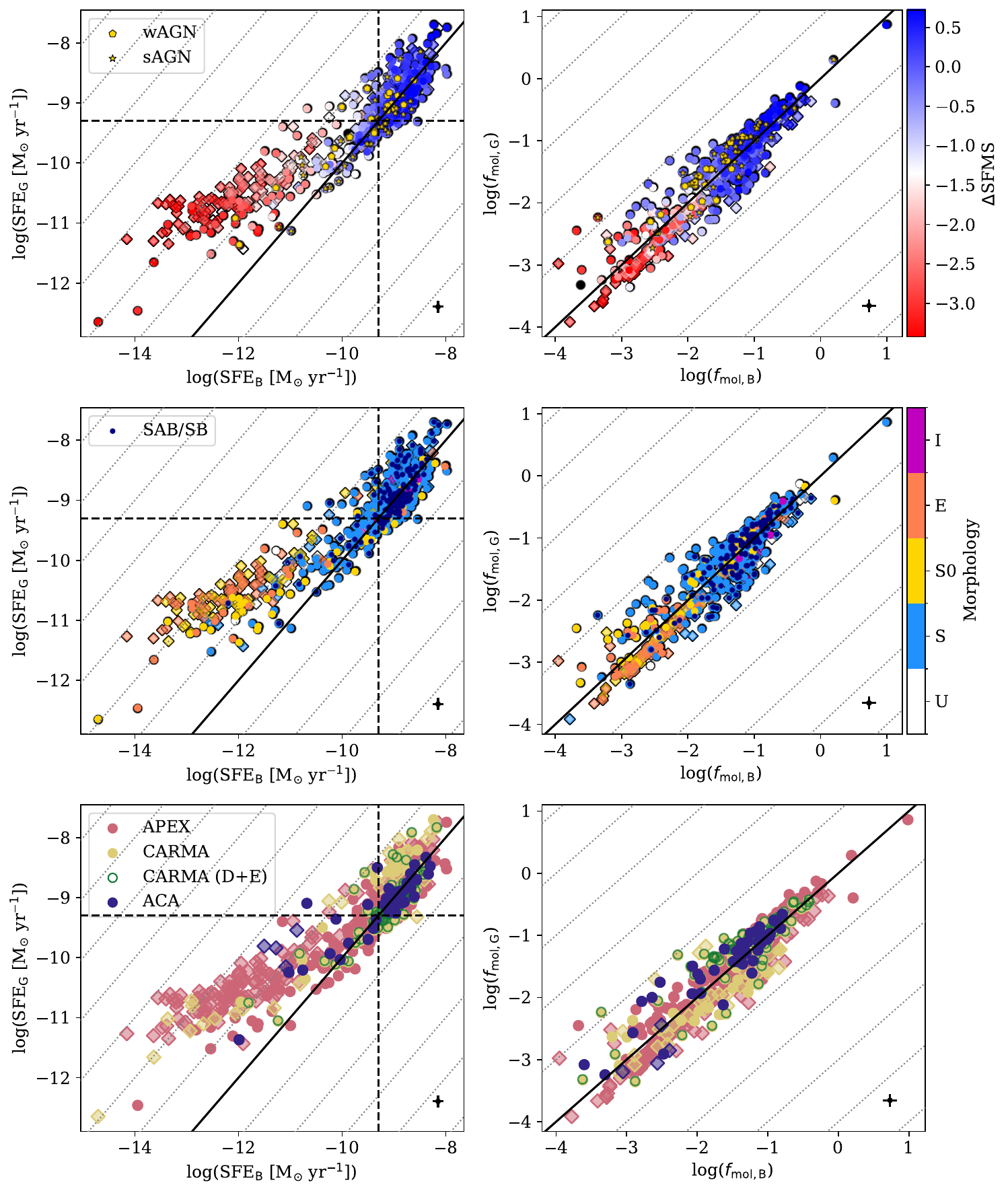}
    \caption{Relationships between global and beam quantities, namely star formation efficiency, SFE, left column, and molecular gas fraction $f_{\rm mol}$, right column, colour-coded by the distance from the SFMS ($\Delta$SFMS, top row), morphology (middle row), and the different samples included in the database (bottom row). In the panels, the black full line shows the 1:1 locus, while grey dotted lines are separated by 1 dex. The black dashed lines in the left column indicate a $\tau_{\rm dep}=2$\, Gyr. In the top row, the yellow pentagons show the positions of the wAGN-hosts in the sample, while the yellow stars show the positions of sAGN-hosts. In the middle row, dark blue dots show the locations of the SAB/SB galaxies. In the panels, the transparent diamonds are used to indicate that a given measurement is obtained with a CO upper limit. While $f_{\rm mol}$ does not appear to vary across the galaxy discs, the SFE in the centre of the retired early-type galaxies in particular is  $\sim1$\, dex lower compared to the SFE across the entire galaxies.}
    \label{F:sfe_fmol_globbeam}
\end{figure*}

The database that we constructed, including both beam and global measurements, allows us to study how the overall behaviours of star formation efficiency (SFE) and molecular gas fraction ($f_{\rm mol}$) compare with their values in the galaxy inner regions (typically within 1\,$R_{\rm e}$). The results of this analysis are reported in Fig.~\ref{F:sfe_fmol_globbeam}, where global versus beam SFE and $f_{\rm mol}$ are shown colour-encoded for $\Delta$SFMS and the galaxy morphology. Generally, both $f_{\rm mol}$ and SFE decrease by several orders of magnitude from spirals to early-type galaxies. However, while the fraction of molecular gas is spatially constant across the galaxies, the inner SFE is lower compared to the global SFE, especially for passive objects.

In particular, the SFE$_{\rm G}$ is slightly shifted towards higher values compared to SFE$_{\rm B}$ but is still well represented by the 1:1 relation in star-forming spirals, the relation between SFE$_{\rm G}$ and SFE$_{\rm B}$ is flatter across retired, E-S0 galaxies (left panels of Fig.~\ref{F:sfe_fmol_globbeam}). In these systems, the inner regions form stars at a rate that is (roughly) 10 times lower than the global value. This trend shift appears to happen for galaxies of about 1.5-2 dex away from the SFMS (e.g. having $\Delta$SFMS$\sim-1.5\div-2$). The flatter relation is less defined for elliptical galaxies (which are mostly non-detected), but it is quite clear for the lenticular galaxies. The same behaviour is not observed in $f_{\rm mol}$ diagrams (right panels of Fig.~\ref{F:sfe_fmol_globbeam}) where $f_{\rm mol, G}$ and $f_{\rm mol, B}$ are consistently scattered (or $\sim1$ dex) across the 1:1 relation from the spiral, star-forming galaxies to the retired, early-type systems. This means that the availability of molecular gas does not significantly vary across the galactic discs. Therefore, the reason for low SFE does not seem to be connected with the scarcity of molecular gas in the inner galactic regions but to a reduced SFR. This is a clear manifestation of the inside-out quenching observed in nearby galaxies \citep[e.g.][]{gonzalez_delgado2016,ellison2018,bluck2019,kalinova2021,kalinova2022}, although with our database, we cannot robustly separate inner and outer galaxy regions. Even from the currently available spatially resolved data, explaining whether variations of $f_{\rm mol}$ or SFE cause inside-out quenching is challenging, mostly due to the limited sample of galaxies observed in CO-lines in the green valley. Indeed, in green valley galaxies, \cite{villanueva2024} measured lower $f_{\rm mol}$ but similar SFE within 1\,$R_{\rm e}$ compared to the star-forming sample; \cite{pan2024} showed instead that within 0.5\,$R_{\rm eff}$ the quenching is primarily attributable to reduce SFE; while \cite{brownson2020} indicate that both reduced SFE and $f_{\rm mol}$ are necessary to turn off the star formation in green valley galaxies. Still, those conclusions are derived from a handful of galaxies.

Additionally, there is a tendency for active galaxies to have roughly equal SFE$_{\rm G}$ and SFE$_{\rm B}$ but slightly higher $f_{\rm mol, G}$ compared to $f_{\rm mol, B}$. While this means that AGNs contribute to reducing the gas in the centre of galaxies (therefore having lower molecular gas fractions in the inner regions than globally), it is hard to tell through integrated measurements only.

Instead, barred galaxies have roughly equal SFE$_{\rm G}$ and SFE$_{\rm B}$, and $f_{\rm mol, G}$ and $f_{\rm mol, B}$. We only notice that some barred galaxies have SFE$_{\rm G}>$SFE$_{\rm B}$; however, this follows the global trend of the spiral galaxies, where several objects have SFE$_{\rm G}>$SFE$_{\rm B}$. This would rule out quenching scenarios related to the bar's presence in our sample. Once again, it is hard to establish this with integrated measurements only.

This analysis also offers the opportunity to compare the SFEs and molecular gas fractions measured across the different sub-samples present in the database (APEX, ACA, CARMA, and CARMA D+E), mimicking Fig.~\ref{F:sfm_sample_det}. We can observe that ACA and CARMA sub-samples are constrained towards high SFEs, typically above $10^{-10}$\, M$_{\odot}$\,yr$^{-1}$, while APEX data extend towards 2 to 4 orders of magnitude lower efficiency measurements (see Fig.~\ref{F:sfe_fmol_globbeam}, bottom panels). Nevertheless, the low SFE tail of the distributions is dominated by non-detections. Given this and that global measurements from APEX data depend upon the aperture correction applied, the behaviour of SFE needs to be cautiously considered. The same is not entirely true for \fmol, where the ACA sample, in particular, covers a fairly similar parameter space as the APEX sample. However, galaxies with $f_{\rm mol}<10^{-2}$ are largely under-represented by the interferometric sub-samples.

\section{Conclusion}\label{S:conclusion}
In this paper, we have presented \iedge, a new database to study galaxy evolution in the local Universe. \iedge\ includes integrated measurements for 643 spiral, elliptical, and lenticular galaxies spanning the full SFR-$M_*$ diagram. Additionally, the database comprises integrated, global quantities, and measurements from the inner regions of the galaxies (within 23.6 arcseconds, or the APEX beam at 230\, GHz, roughly corresponding to 1\,$R_{\rm e}$ for most of the galaxies, e.g. beam measurements). Optical wavelength quantities are derived from the CALIFA survey, while CO-line information is derived from APEX, CARMA, and ACA observations collected by the EDGE-CALIFA collaboration. As the database is built from data collected from IFU, radio single-dish and interferometers, quantities needed to be properly homogenised and organised as follows:

\begin{itemize}
    \item CALIFA global quantities (such as SFRs and stellar masses) are calculated by integrating across the maps, while beam quantities are measured by integrating tapered maps. The tapering map is obtained using a 2D Gaussian with unitary amplitude and FWHM=23.6 arcsec.
    \item APEX global CO luminosities are derived using an aperture correction method that uses 12\,$\mu$m maps from WISE W3 data collected by z0MGS dataset, while APEX beam quantities are calculated directly from the spectra.
    \item CO global quantities from interferometers are assumed from their respective publications (\citealt{bolatto2017}, for CARMA data; and \citealt{villanueva2024}, for ACA data), while beam measurements are inferred from the single spectrum at the centre of the datacube, smoothed at the spatial and spectral resolutions of the APEX data.
    \item Other quantities, such as the CO-to-H$_2$ conversion factors and the CO(2-1)-to-CO(1-0) ratios are modelled from the IFU data products.
    \item The final dataset includes 454 APEX, 47 ACA, 100 CARMA (D+E), and 42 CARMA (E) galaxies.
\end{itemize}

We used the \iedge\ to define scaling relations between quantities fundamental to study galaxy evolution: the SFR-$M_*$, SFR-$M_{\rm mol}$, and $M_{\rm mol}-M_*$ relations across morphological types and the variation of star formation efficiency (SFE) and molecular gas fraction ($f_{\rm mol}$) from the inner regions to the galaxies to the full extend of their discs. We observed that:

\begin{itemize}
    \item The three diagrams are bi-modal, with spiral, star-forming, and early-type passive galaxies distributed across well-separated relationships. In particular, while spirals generally organise across $\tau_{\rm dep}\sim2$\,Gyr and $f_{\rm mol}\sim0.1$, early types show average $\tau_{\rm dep}>50$\,Gyr and $f_{\rm mol}<0.01$.
    \item While molecular gas fractions appear to be largely constant across the galaxies, the SFE in the inner regions of the galaxies is $\sim2$ dex lower than the global value for galaxies with $\Delta$SFMS$<-1.5$. This is a clear manifestation of the inside-out quenching that requires not only lower molecular gas mass availabilities but also strongly reduced SFE in the centre of the galaxies to move objects from the SFMS to the retired region. Nevertheless, as the bending observed in the SFE diagram is largely driven by APEX data, for which an aperture correction is applied to extrapolate global molecular gas masses, its significance should be interpreted with caution, as it may be influenced by the correction rather than solely reflecting an intrinsic trend.
    \item Our data do not show significant differences between active and non-active galaxies, or barred and un-barred galaxies, but only a small indication that beam $f_{\rm mol}$ are slightly lower than global $f_{\rm mol}$ in active galaxies, indicating possible reduced molecular gas availability in the centre of the galaxies due to AGN feedback.
\end{itemize}

Currently, \iedge\ is the most extended, homogeneous, and unbiased collection of integrated galaxy quantities to date. The combination of optical IFU and CO-based measurements makes it particularly useful to constrain the mechanisms that drive galaxy evolution, and in particular, the star formation quenching in the local Universe. 

\section*{Data availability}
The iEDGE (described in Table~\ref{T:database_entry}) is available in electronic form on Zenodo at \url{https://doi.org/10.5281/zenodo.15822433}.
The full figure of Appendix B is available on Zenodo at \url{https://doi.org/10.5281/zenodo.15648223}.

\begin{acknowledgements}
This paper is dedicated to the memory of Karl Menten, who made this work (and many others) possible with his baby-telescope, APEX. The authors thank the anonymous referee for the precious comments that significantly improved the paper's quality. DC thanks Kathryn Kreckel for the useful discussions about metallicity estimators. DC, ZB, and FB gratefully acknowledge the Collaborative Research
Center 1601 (SFB 1601 sub-project B3) funded by the Deutsche Forschungsgemeinschaft (DFG, German Research Foundation) –
500700252. DC and AW acknowledge support by the \emph{Deut\-sche For\-schungs\-ge\-mein\-schaft, DFG\/} project number SFB956-A3. SFS thanks the PAPIIT-DGAPA AG100622 project and CONACYT grant CF19-39578. This work was supported by UNAM PASPA – DGAPA. ER acknowledges the support of the Natural Sciences and Engineering Research Council of Canada (NSERC), funding reference number RGPIN-2017-03987. ADB, TW, LB, SV, and RCL acknowledge support from the National Science Foundation (NSF) through the collaborative research award AST-1615960. KDF acknowledges support from NSF grant AAG 23-07440. TW and YC acknowledge support from the NSF through grant AST-1616199. JBB acknowledges support from the grant IA-100420 (PAPIIT-DGAPA, UNAM). JBB acknowledges funding from the grant IA-101522 (DGAPA-PAPIIT, UNAM) and support from the DGAPA-PASPA 2025 fellowship (UNAM). VV acknowledges support from the ALMA-ANID Postdoctoral Fellowship under the award ASTRO21-0062. RCL acknowledges partial support for this work from the NSF under award AST-2102625. This research made use of Astropy,\footnote{http://www.astropy.org} a community-developed core Python package for Astronomy \citep{astropy2013, astropy2018}; matplotlib \citep{matplotlib2007}; and numpy and scipy \citep{scipy2020}. Support for CARMA construction was derived from the Gordon and Betty Moore Foundation, the Eileen and Kenneth Norris Foundation, the Caltech Associates, the states of California, Illinois, and Maryland, and the NSF. Funding for CARMA development and operations was supported by NSF and the CARMA partner universities. The authors acknowledge the usage of the HyperLeda database (\url{http://leda.univ-lyon1.fr}), the NASA/IPAC Extragalactic Database (\url{http://ned.ipac.caltech.edu}), and the NASA/IPAC Infrared Science Archive (\url{https://irsa.ipac.caltech.edu/frontpage/}).
\end{acknowledgements}

\footnotesize{
\bibliographystyle{aa}
\bibliography{cold}
}

\appendix

\section{CO(1-0)-to-H$_2$ conversion factor dependences on gas-phase metallicity and stellar mass surface density}
\label{A:aco}

In this work, we adopt a CO(1-0)-to-H$_2$ conversion factor (\aco) modelled following the prescription of \cite{bolatto2013}, which accounts for dependences on gas-phase metallicity and total mass surface density (here assumed to be equivalent to the stellar surface density). Among these parameters, gas-phase metallicity is supposed to play a dominant role, as \aco\ exhibits an exponential dependence on it (see equation \ref{E:aco}). We compute gas-phase metallicity on a spaxel-by-spaxel basis using the O3N2 method \citep{pettini_pagel2004}, which is applicable only in regions where ionisation is driven by star formation. To extend the metallicity estimation to regions where ionisation is dominated by old stars or AGNs, we construct a radial metallicity map based on the mass-metallicity relation (MZR) from \cite{sanchez2017}.

\begin{figure}
\centering
\includegraphics[width = 0.4\paperwidth, keepaspectratio]{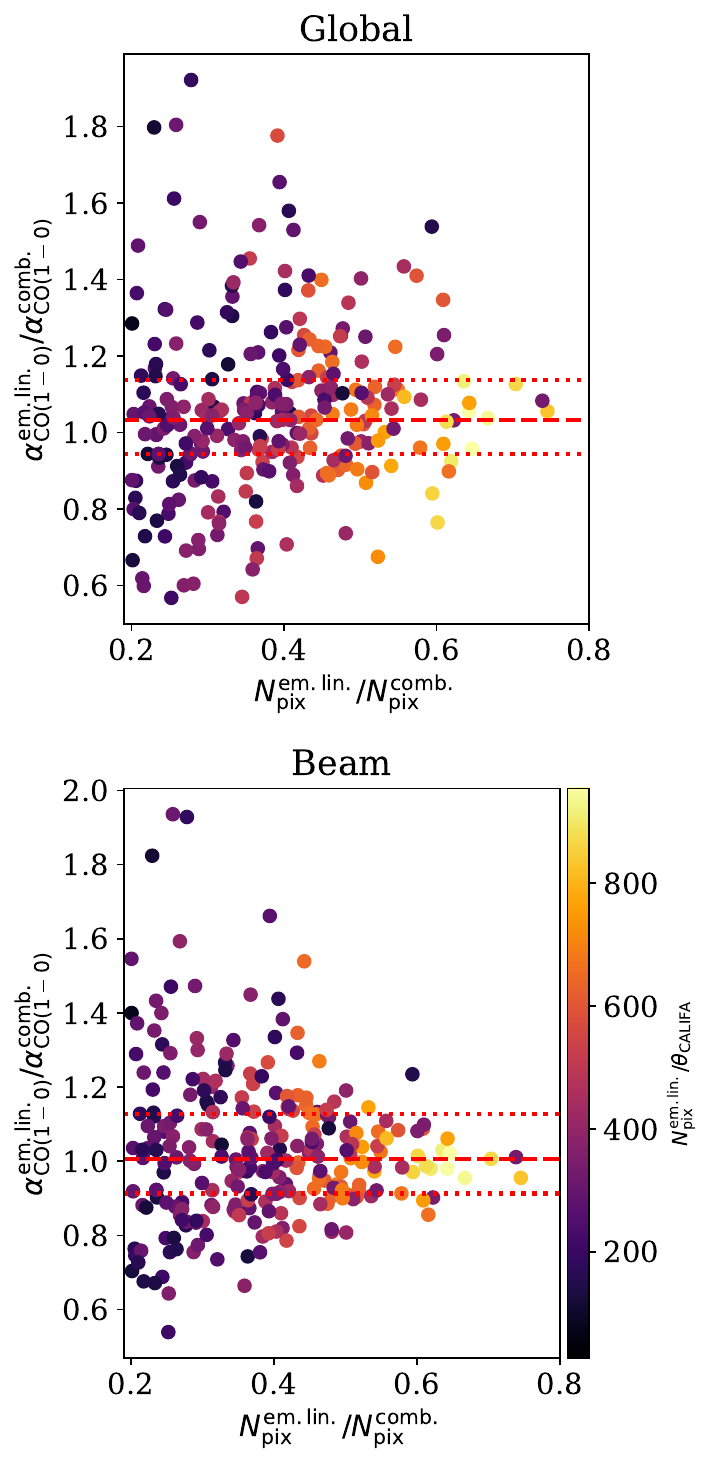}
\caption{Ratio of the \aco\ calculated from the gas-phase metallicity inferred only from nebular emission lines ($\alpha^{\rm em. lin.}_{\rm CO(1-0)}$) to the combined map ($\alpha^{\rm comb.}_{\rm CO(1-0)}$) vs the ratio of the number of pixels within the emission lines-map to the number of pixels within the combined map. The colour bar shows the number of resolution elements ($\theta_{\rm PSF}$) that constitute the gas-phase metallicity map from the nebular emission lines. The red horizontal lines (from  bottom to  top) indicate the $25{\rm th}$, $50{\rm th}$, and $75{\rm th}$ percentiles of the \aco\ ratio distributions. Additionally, \aco\ values from the left panel come from global measurements, while on the right panels beam measurements are displayed.}
\label{F:aco_elcomb}
\end{figure}

Figure~\ref{F:aco_elcomb} compares the \aco\ values obtained from this combined metallicity map with those derived solely from emission-line metallicities. For a robust comparison, we consider only emission-line metallicity maps that contain at least 20\% of the spaxels of the combined metallicity map. This analysis reveals that \aco\ values inferred from the combined metallicity map can differ by up to a factor of 2 from those based solely on emission-line metallicities. The largest discrepancies arise typically in cases where the fraction of spaxels with detectable emission lines is small relative to the total number of spaxels in the combined map. The ratio tends to unity for galaxies where the number of spaxels within the emission line metallicity map becomes equivalent to the combined metallicity map. This is particularly true in the case of the beam \aco\ measurements, as nebular emission lines are typically detected in the central regions of the galaxies. Nonetheless, the observed discrepancies remain within the typical uncertainty range associated with $\alpha_{\rm CO(1-0)}$ determinations (approximately 0.3 dex; \citealt{bolatto2013}), indicating that the choice between combined and emission line-only metallicity maps does not introduce a significant systematic bias in the resulting molecular gas mass estimates.

\begin{figure*}
\centering
\includegraphics[width = 0.9\paperwidth, keepaspectratio]{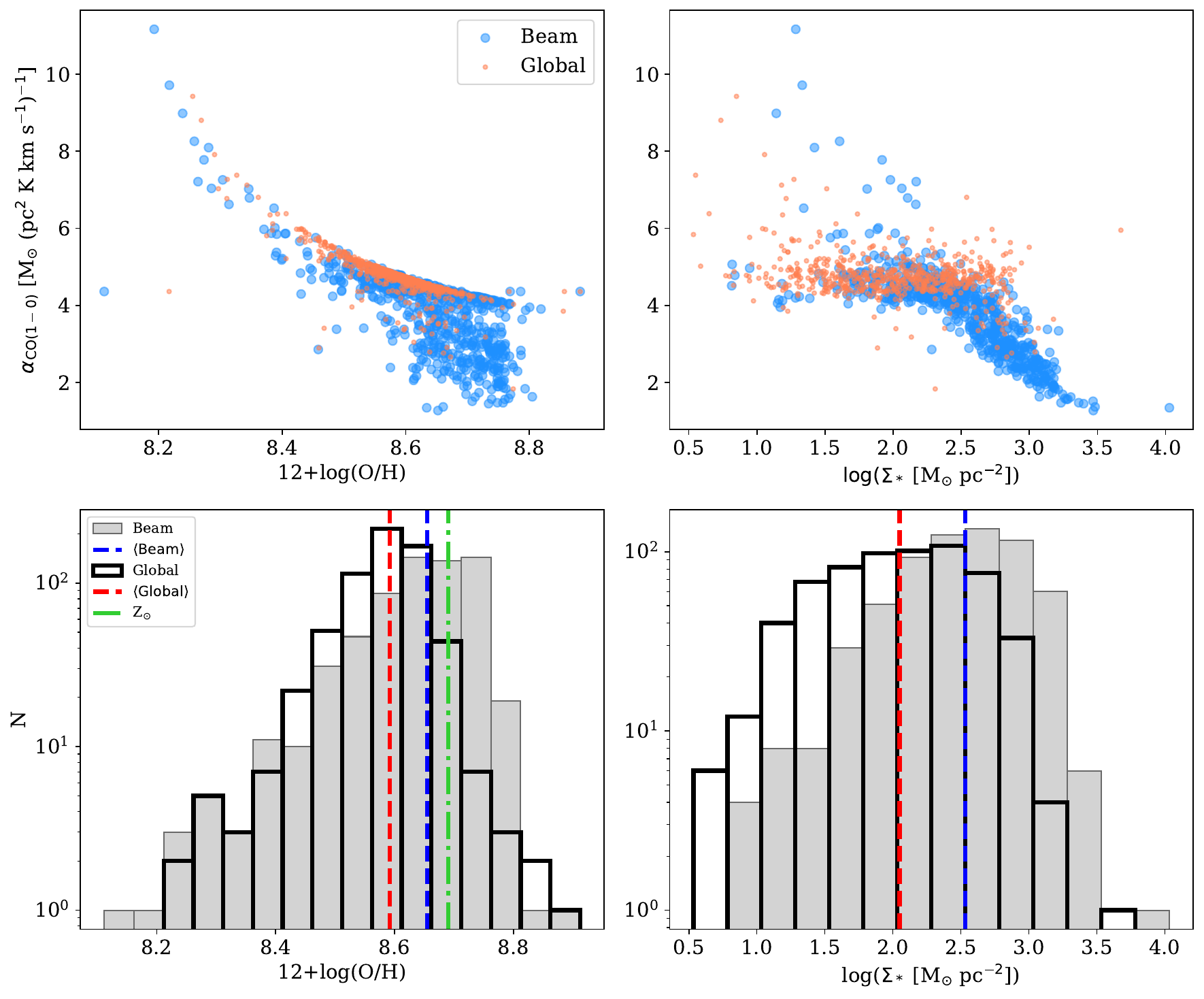}
\caption{\emph{Upper row:} Relation between beam (blue circles) and global (red dots) \aco\ measurements on gas-phase metallicity (12+log(O/H), left panel) and stellar mass surface density ($\Sigma_*$ right panel) calculated within the same regions. \emph{Bottom row:} Distributions of metallicity and stellar mass surface density. The histograms show median metallicity and integrated $\Sigma_*$ calculated across the full maps (global, empty black histograms) and tapered maps (beam, grey histograms). The red (blue) dashed lines indicate the median of the  global  (beam) quantities, while the green dash-dotted line (left panel) shows the solar metallicity (8.69 by \citealt{allende_prieto2001}).}
\label{F:aco_met_smstars}
\end{figure*}

To evaluate the relative dependence of \aco\ on metallicity and stellar mass surface density, Fig.~\ref{F:aco_met_smstars} (upper row) presents the correlation of global and beam \aco\ measurements with these two parameters, both derived from the same galactic regions. We find that both global and beam \aco\ measurements exhibit a strong anticorrelation with gas-phase metallicity. However, while global \aco\ remains largely insensitive to $\Sigma_*^{\rm G}$, beam \aco\ reaches systematically lower values due to a significant dependence on $\Sigma_*^{\rm B}$.

The lower row of Fig.\ref{F:aco_met_smstars}, analogous to Fig.\ref{F:r21_aco_histo}, illustrates the distributions of global and beam gas-phase metallicity and stellar mass surface density. Here, beam stellar mass surface densities are calculated considering the beam stellar mass divided by the physical area defined by the APEX beam, while the global stellar mass surface densities are given by the global stellar mass divided by the physical area encompassed by $2\, R_{\rm e}$, which is the typical coverage of the galaxies by the CALIFA IFU. Beam-evaluated metallicity measurements are systematically shifted towards higher values compared to their global counterparts, consistent with the lower beam \aco\ values observed in Section~\ref{SSS:sample_stat_r21_aco}. The distributions of $\Sigma_*$ measurements exhibit a similar trend; however, this appears to have a minimal impact on the final global \aco\ values. These comparisons confirm that the adopted \aco\ prescription is primarily governed by its dependence on gas-phase metallicity, while variations in stellar mass surface density, although affecting beam-scale estimates, do not significantly alter the resulting global values, thus validating the robustness of the chosen parametrisation.

\section{Menagerie of galaxies and spectra in the \iedge}
\label{A:gal_spectra}
In this section, we collected the discrete H$\alpha$ flux maps of the galaxies together with the CO spectra observed by APEX, CARMA and/or ACA. In the following, H$\alpha$ flux maps are shown in the first column of the figure, where the colour maps are based on the galaxy morphology: blue stretches for spirals, red stretches for elliptical, yellow stretches for lenticular, and grey stretches for irregular or galaxies with undefined morphology. CO spectra from APEX, CARMA, and ACA (if available) are shown in the right columns. In particular, CARMA D+E configuration data are shown in yellow, while CARMA E-only configuration data are in green. The full figure is available on Zenodo at \href{https://zenodo.org/records/15387356?token=eyJhbGciOiJIUzUxMiJ9.eyJpZCI6ImE3OTdkODQ1LWViNzMtNDZhMS05ODA5LWQxYTgzMTA5NzQ5MSIsImRhdGEiOnt9LCJyYW5kb20iOiJlZDUwMjE1YThiNmY5NWY5ZWI1OTQ1M2MzYzAyYzc1OSJ9.NDP6NAEb-dW0fJ66-gbem_pw2DqeA4hpKa6XxhiAadjuMDgOw3JEfO5MVARo3kh8JPF5_lyBbZTisEpuscuP8Q}{https://zenodo.org/records/15387356}.

\section{Content of the \iedge\ current data release}

\begin{table*}
\label{T:database_entry}
\caption{Quantities included in the current data release of the \iedge.}
\setlength{\tabcolsep}{2pt}
\centering
\begin{tabular}{rcl}
\hline
\hline
Database entry & Unit & Description \\
\hline
(1) & (2) & (3)\\
\hline
CALIFA\_name & & Name of the galaxy as indicated in the CALIFA extended DR3 \\
Alias & & Alternative name of the galaxy \\
RA & deg & Right ascension of the galaxy centre \\
DEC & deg & Declination of the galaxy centre \\
Vlsr & km s$^{-1}$ & LSR velocity of the galaxy, derived from the stellar redshift \\
zstar & & Stellar redshift of the galaxy \\
Distance & Mpc & Distance to the galaxy \\
Incl & deg & Galaxy inclination \\
PA & deg & Galaxy position angle \\
Re\_as & arcsec & Galaxy effective radius \\
\emph{Type}\_med\_EWHa & $\AA$ & Median H$\alpha$ equivalent width \\
\emph{Type}\_Mstar & M$_{\odot}$ & Stellar mass \\
\emph{Type}\_SFR & M$_{\odot}$\,yr$^{-1}$ & Star formation rate derived using Balmer decrement \\
\emph{Type}\_sSFR & yr$^{-1}$ & Specific star formation rate \\
Err\_\emph{Type}\_Mstar & M$_{\odot}$ & Uncertainty on the stellar mass \\
Err\_\emph{Type}\_SFR & M$_{\odot}$\,yr$^{-1}$ & Uncertainty on the star formation rate derived using  Balmer decrement \\
Err\_\emph{Type}\_sSFR & yr$^{-1}$ & Uncertainty on the specific star formation rate \\
\emph{Type}\_Gas\_Met &  & Gas-phase metallicity derived via the O3N2 method \\
\emph{Type}\_alpha\_co10 & M$_{\odot}$\,(K\,km\,s$^{-1}$)$^{-1}$ & CO(1-0)-to-H$_2$ conversion factor derived following \cite{bolatto2013} \\
\emph{Type}\_R21 & & CO(2-1)/CO(1-0) flux ratio derived from the star formation rate surface density \\
\emph{Observatory}\_\emph{Type}\_k2jypb & Jy\,beam$^{-1}$\,K$^{-1}$ & Jy/beam per Kelvin \\
\emph{Observatory}\_\emph{Type}\_eta\_mb &  & Main beam efficiency \\
\emph{Observatory}\_\emph{Type}\_RMS & K & CO noise level (standard deviation of CO spectrum, $\sigma_{\rm RMS}$) \\
\emph{Observatory}\_\emph{Type}\_Tpeak & K & Peak $T_{\rm mb}$ across the CO spectrum \\
\emph{Observatory}\_\emph{Type}\_SNR & & Signal-to-noise ratio of the CO observation \\
\emph{Observatory}\_\emph{Type}\_W50\_mom2 & & Full-width half maximum of the CO spectrum calculated via moment 2nd \\
\emph{Observatory}\_\emph{Type}\_eps\_obs & K\,km\,s$^{-1}$ & Statistical error for the CO flux \\
\emph{Observatory}\_\emph{Type}\_Fco & Jy\,beam$^{-1}$\,km\,s$^{-1}$ & CO flux \\
\emph{Observatory}\_\emph{Type}\_Lco & K\,km\,s$^{-1}$\,pc$^2$ & CO luminosity \\
\emph{Observatory}\_\emph{Type}\_Mmol & M$_{\odot}$ & Molecular gas mass \\
\emph{Observatory}\_\emph{Type}\_fmol & & Molecular gas mass to stellar mass ratio \\
\emph{Observatory}\_\emph{Type}\_SFE & yr$^{-1}$ & Star formation efficiency \\
Final\_\emph{Type}\_Observatory & & Observatory of the CO measurement included in the consolidated sample \\
Final\_\emph{Type}\_SNR & & Signal-to-noise ratio of the CO observation from the consolidated sample \\
Final\_\emph{Type}\_Mmol & M$_{\odot}$ & Molecular gas mass from the consolidated sample \\
Final\_\emph{Type}\_fmol & & Molecular gas mass fraction from the consolidated sample \\
Final\_\emph{Type}\_SFE & yr$^{-1}$ & Star formation efficiency from the consolidated sample \\
\emph{Observatory}\_Err\_\emph{Type}\_Fco & Jy\,beam$^{-1}$\,km\,s$^{-1}$ & Uncertainty on the CO flux \\
\emph{Observatory}\_Err\_\emph{Type}\_Lco & K\,km\,s$^{-1}$\,pc$^2$ & Uncertainty on the CO luminosity \\
\emph{Observatory}\_Err\_\emph{Type}\_Mmol & M$_{\odot}$ & Uncertainty on the molecular gas mass \\
\emph{Observatory}\_Err\_\emph{Type}\_fmol & & Uncertainty on the molecular gas mass to stellar mass ratio \\
\emph{Observatory}\_Err\_\emph{Type}\_SFE & yr$^{-1}$ & Uncertainty on the star formation efficiency \\
Final\_Err\_\emph{Type}\_Mmol & M$_{\odot}$ & Uncertainty on the molecular gas mass from the consolidated sample \\
Final\_Err\_\emph{Type}\_fmol & & Uncertainty on the molecular gas fraction from the consolidated sample \\
Final\_Err\_\emph{Type}\_SFE & yr$^{-1}$ & Uncertainty on the star formation efficiency from the consolidated sample \\
\emph{Type}\_L12um\_w3\_7p5 & K\,km\,s$^{-1}$\,pc$^2$ & 12\,$\mu$m luminosity from WISE W3 at 7.5 arcsec resolution \\
Err\_\emph{Type}\_L12um\_w3\_7p5 & K\,km\,s$^{-1}$\,pc$^2$ & Uncertainty on the 12\,$\mu$m luminosity from WISE W3 at 7.5 arcsec resolution \\
CARMA\_conf & & Array configuration of the CARMA observations\\
Morphology & & Galaxy morphology obtained from Hyperleda \\
\hline
\end{tabular}
\tablefoot{\emph{Type} considers global (Glob) and beam (Beam) quantities included in the database. \emph{Observatory} indicates the facility from which the CO observation was performed (APEX, CARMA, or ACA). In the case a given source was not observed with the facility, the quantity assumes a NaN entry. Quantities measured from non-detected CO observation (with S/N<3) are included as upper limits. For the morphology, we simply separated the galaxies as spiral (S), elliptical (E), lenticular (S0), and irregular (I), without considering the sub-classes of spirals (except intermediate spirals, SABs; and barred spirals, SBs) and elliptical morphologies. Galaxies for which the morphology is not provided are indicated with `U'.}
\end{table*}

\end{document}